\documentclass[12pt,preprint]{aastex}
\shorttitle{Wind Variations of Cyg X-1}
\shortauthors{Gies et al.}
\begin{document}


\title{Stellar Wind Variations During the X-ray \\ 
High and Low States of Cygnus X-1 \altaffilmark{1,2}}

\altaffiltext{1}{Based on observations with the NASA/ESA Hubble 
Space Telescope obtained at the Space Telescope Science Institute,
which is operated by the Association of Universities for Research
in Astronomy, Incorporated, under NASA contract NAS5-26555.
These observations are associated with programs GO-9646 and GO-9840.}

\altaffiltext{2}{Based on data obtained at the David Dunlap Observatory, 
University of Toronto.} 

\author{D. R. Gies\altaffilmark{3,4}, C. T. Bolton\altaffilmark{5}, 
R. M. Blake\altaffilmark{5}, S. M. Caballero-Nieves\altaffilmark{3}, 
D. M. Crenshaw\altaffilmark{3}, \\ P. Hadrava\altaffilmark{6}, 
A. Herrero\altaffilmark{7}, T. C. Hillwig\altaffilmark{8}, 
S. B. Howell\altaffilmark{9}, W. Huang\altaffilmark{4,10},
L. Kaper\altaffilmark{11}, \\ P. Koubsk\'{y}\altaffilmark{6}, 
and M. V. McSwain\altaffilmark{12,13}}

\altaffiltext{3}{Center for High Angular Resolution Astronomy,
Department of Physics and Astronomy, 
Georgia State University, P. O. Box 4106, Atlanta, GA  30302-4106;
gies@chara.gsu.edu, scaballero@chara.gsu.edu, crenshaw@chara.gsu.edu}
 
\altaffiltext{4}{Visiting Astronomer, Kitt Peak National Observatory,
National Optical Astronomy Observatory, operated by the Association
of Universities for Research in Astronomy, Inc., under contract with
the National Science Foundation.}

\altaffiltext{5}{David Dunlap Observatory, University of Toronto, 
P. O. Box 360, Richmond Hill, Ontario, L4C 4Y6, Canada;
bolton@astro.utoronto.ca}

\altaffiltext{6}{Astronomical Institute, Academy of Sciences of the Czech Republic, 
Fri\v{c}ova 298, CZ-251 65 Ond\v{r}ejov, Czech Republic; had@sunstel.asu.cas.cz, 
koubsky@sunstel.asu.cas.cz}

\altaffiltext{7}{Instituto de Astrof\'{i}sica de Canarias, 38200, La Laguna, Tenerife, Spain; 
Departamento de Astrof\'{i}sica, Universidad de La Laguna, 
Avda. Astrof\'{i}sico Francisco S\'{a}nchez, s/n, 38071 La Laguna, Spain;
ahd@ll.iac.es}

\altaffiltext{8}{Department of Physics and Astronomy, Valparaiso University, 
Valparaiso, IN 46383; todd.hillwig@valpo.edu}

\altaffiltext{9}{WIYN Observatory and National Optical Astronomy Observatory, 
P. O. Box 26732, 950 N. Cherry Ave., Tucson, AZ 85719; howell@noao.edu} 

\altaffiltext{10}{Department of Astronomy, California Institute of Technology, MC 105-24,
Pasadena, CA 91125; wenjin@astro.caltech.edu}

\altaffiltext{11}{Astronomical Institute Anton Pannekoek, Universiteit van Amsterdam, 
Kruislaan 403, 1098-SJ Amsterdam, The Netherlands; lexk@science.uva.nl}

\altaffiltext{12}{Department of Physics, Lehigh University, 
16 Memorial Drive East, Bethlehem PA 18015; 
mcswain@lehigh.edu}

\altaffiltext{13}{Guest investigator, Dominion Astrophysical Observatory, 
Herzberg Institute of Astrophysics, National Research Council of Canada}

\slugcomment{ApJ, in press}
\paperid{72368}


\begin{abstract}
We present results from {\it Hubble Space Telescope} ultraviolet 
spectroscopy of the massive X-ray and black hole binary system, 
HD~226868 = Cyg X-1.  The spectra were obtained at both 
orbital conjunction phases in two separate runs in 2002 and 2003
when the system was in the X-ray high/soft state.  
The UV stellar wind lines suffer large reductions in 
absorption strength when the black hole is in the foreground   
due to the X-ray ionization of the wind ions.  
A comparison of the {\it HST} spectra with 
archival, low resolution spectra from the 
{\it International Ultraviolet Explorer Satellite} shows that 
similar photoionization effects occur in both the 
X-ray high/soft and low/hard states.   We constructed 
model UV wind line profiles assuming that X-ray ionization 
occurs everywhere in the wind except the zone where the supergiant 
blocks the X-ray flux.  The good match between the observed 
and model profiles indicates that the wind ionization extends 
to near to the hemisphere of the supergiant facing the X-ray source.  
We also present contemporaneous spectroscopy of the H$\alpha$ emission
that forms in the high density gas at the base of the supergiant's wind
and the \ion{He}{2} $\lambda4686$ emission that originates in the 
dense, focused wind gas between the stars.  
The H$\alpha$ emission strength is generally lower in 
the high/soft state compared to the low/hard state, but the 
\ion{He}{2} $\lambda4686$ emission is relatively constant 
between X-ray states.  The results suggest that mass transfer 
in Cyg~X-1 is dominated by the focused wind flow that peaks 
along the axis joining the stars and that the stellar wind 
contribution from the remainder of the hemisphere facing the 
X-ray source is shut down by X-ray photoionization effects 
(in both X-ray states).  The strong stellar wind from the 
shadowed side of the supergiant will stall when Coriolis 
deflection brings the gas into the region of X-ray illumination. 
This stalled gas component may be overtaken by the orbital 
motion of the black hole and act to inhibit 
accretion from the focused wind.  The variations in the 
strength of the shadow wind component may then lead to 
accretion rate changes that ultimately determine the 
X-ray state.  
\end{abstract}



\setcounter{footnote}{13}

\section{Introduction}                              

The massive X-ray binary Cygnus X-1 is a seminal target in the 
study of gas dynamics in the vicinity of a stellar mass black hole. 
Its X-ray luminosity and energetic jets \citep{gal05} are
powered by gas accretion from the nearby companion star 
HD~226868 (O9.7~Iab; \citealt{wal73}) 
in a spectroscopic binary with a 5.6 day orbital period. 
There are several ways in which mass transfer from the supergiant 
to the black hole may occur in this system \citep{kap98}.  
The O-supergiant, like other massive and luminous stars, has a 
strong radiatively driven wind that may be partially accreted 
through the gravitational force of the black hole.  The supergiant
is large and is probably close to filling its critical Roche 
surface \citep{gie86a,her95}, so a gas stream through the inner
L1 point may also be present.  The actual gas flow in the direction
of the black hole is probably intermediate between a spherically 
symmetric wind and a Roche lobe overflow stream, and there is 
evidence that the flow is best described as a focused wind 
\citep{fri82,gie86b,gie03,mil05}.  The gas ions responsible for 
accelerating the wind may become ionized in the presence of a
strong X-ray source, leading to a lower velocity, ``stalled'' wind 
\citep{blo90,ste91}.  In situations of very high X-ray flux, 
photoionization may extend so close to the supergiant's photosphere 
that the wind never reaches the stellar escape velocity and 
thus ceases to become an X-ray accretion source \citep{day93,blo94}.
However, such a high X-ray flux may heat the outer gas layers 
to temperatures where the thermal velocities exceed the escape 
velocity to create a thermal wind that may fuel black hole accretion. 

Important clues about the mass transfer process come from the 
temporal variations of the observed X-ray flux.  Cyg~X-1 is generally 
observed in either a low flux/hard spectrum state, with an X-ray 
spectrum that is relatively flat, or a high flux/soft spectrum state 
with a steeper power-law spectrum \citep{sha06}.  The gamma-ray 
portion of the spectrum is also elevated during the high/soft state
\citep{mcc02}.  The high/soft state usually lasts for periods of 
days to months, and the fraction of time observed in the high/soft 
state has increased from $10\%$ in 1996--2000 to $34\%$ since 
early 2000 \citep{wil06}.   This increase may be related to 
an overall increase in the supergiant's radius in the period 
from 1997 to 2003 -- 2004 that is suggested by changes in the 
long term optical light curve \citep{kar06a}.   The system 
sometimes experiences so-called failed-state transitions, when it starts 
to increase in flux, but then stops at an intermediate state and returns
to the low/hard state.  All these transitions probably reflect changes 
in the inner truncation radius of the accretion disk surrounding the 
black hole that are caused by a variable accretion rate (largest 
when the system is in the high/soft state; \citealt{don02,mcc06}).  
Thus, the temporal variations in the X-ray state offer us the means
to compare the black hole accretion processes with observational 
signatures related to mass transfer. 

The H$\alpha$ emission formed in the high density gas at the 
base of the stellar wind is an important diagnostic 
of the mass loss rate in massive stars \citep{pul96,mar05}.  
The H$\alpha$ emission variations in HD~226868 over the 
last few years are documented in independent spectroscopic 
investigations by \citet{gie03} and \citet*{tar03}.  Both of these
studies concluded that the H$\alpha$ emission appears 
strongest when the system is in the low/hard X-ray state, 
while a range of weak to moderate emission strengths are 
observed during the high/soft states.  This is a surprising result, 
since taken at face value, strong emission is associated with  
a large wind mass loss rate, and the simplest expectation that 
the X-ray accretion flux increases with mass loss rate is, in fact,
not observed.  There are several possible explanations: 
(1) A denser wind may be more opaque to X-rays.  However, 
this seems unlikely because the observed inverse relation 
between H$\alpha$ emission strength and X-ray flux 
is observed at all orbital phases, not just when 
the supergiant and its wind are in the foreground. 
(2) The X-ray source may photoionize and heat the gas responsible 
for the H$\alpha$ emission, so that a larger X-ray flux leads to 
a decrease in H$\alpha$ strength.  This clearly occurs at some 
level, but both \citet{gie03} and \citet{tar03} argue that portions
of the wind shaded from the ionizing flux also display significant 
temporal variations.  (3) Changes in the X-ray flux will lead to 
variations in the ionized volume of gas surrounding the black hole,
and consequently, the total acceleration of the wind in the direction 
towards the black hole will vary with the distance traveled before the atoms
responsible for line-driving are ionized.  Thus, a stronger, denser 
wind might reach a faster speed before ionization,  and since the 
Bondi-Hoyle accretion rate varies as $\sim v^{-4}$, the gas captured
by the black hole (and the associated X-ray flux) declines.  

This last process can be tested through direct study of the 
degree of wind ionization observed in the ultraviolet P~Cygni lines 
formed in the supersonic part of the wind outflow.   When the 
system is observed with the ionization region in the foreground, 
the absorption cores of these P~Cygni lines will be truncated 
at a blueshift corresponding to the highest projected speed before 
encountering the ionization zone, the so-called Hatchett-McCray 
effect \citep{hat77}.  The binary is so faint in the ultraviolet 
that high dispersion spectra were very difficult to 
obtain with the {\it International Ultraviolet Explorer (IUE)} 
satellite \citep{dav83}, but a good series of observations were 
obtained with {\it IUE} at lower spectral resolution that clearly indicate 
the weakening of the wind lines when the black hole is in front
\citep*{tre80,vlo01}.  Most of these spectra were obtained in 
the low/hard X-ray state (\S5), when according to the varying 
wind strength model the mass loss rate is higher and the wind
is less ionized and faster.  
  
We embarked on a new program of high S/N, high dispersion UV 
spectroscopy to test this hypothesis with the {\it Hubble 
Space Telescope} Space Telescope Imaging Spectrograph (STIS).  
We obtained observations at the two orbital conjunction phases
in both 2002 and 2003.  These sets of observations were both 
made during the rare high/soft X-ray state, and planned observations
during the low/hard state were unfortunately scuttled by the 
STIS electronics failure in 2004.  However, we can rebin the 
high quality {\it HST} spectra made during the high/soft state 
to the lower resolution of the {\it IUE} archival spectra 
(mostly low/hard state) in order to test whether or not 
the wind ionization state does in fact differ significantly 
between states.   We describe a program of supporting
optical spectroscopy we have obtained to check the orbital 
phase (\S2) and wind strength (\S3) at the times
of the {\it HST} observations.  We compare the H$\alpha$ 
measurements with the contemporaneous X-ray light curve 
recorded with the {\it Rossi X-ray Timing Explorer} All-Sky Monitor 
instrument \citep{lev96} and confirm our earlier result 
showing how H$\alpha$ tends to strengthen as the X-ray flux declines (\S3).
We then describe the observed variations in the main UV wind
lines and present a simple ``shadow wind'' model for the profiles (\S4). 
We compare the variations observed in the {\it HST} and {\it IUE}
spectra in \S5, and then reassess the question of the mass transfer 
process in \S6.  We will discuss the photospheric features 
in the UV spectra in a forthcoming paper (Caballero Nieves et al., 
in preparation). 


\section{Observations and Orbital Ephemeris}     

The {\it HST} STIS spectra were obtained with the first-order 
G140M grating in a series of subexposures at different grating 
tilts in order to record the UV spectrum over the full available
range (1150 to 1740 \AA ).  Two full sets were made near 
each orbital conjunction phase in runs in both 2002 and 2003. 
All the spectra were reduced using the IDL software developed at 
NASA Goddard Space Flight Center for the STIS Instrument Definition Team. 
The spectra were rebinned on a $\log \lambda$, heliocentric 
wavelength scale to a spectral resolution of $R=10000$ and 
rectified to a pseudo-continuum based upon the flux in 
relatively line-free regions.  

The space observations were 
supported by contemporaneous, ground-based observations 
of the red spectrum in the vicinity of H$\alpha$.   These 125 
spectra were made with the University of Toronto David Dunlap 
Observatory 1.88~m telescope, NOAO Kitt Peak National Observatory 
0.9~m coud\'{e} feed telescope, Herzberg Institute of Astrophysics 
Dominion Astrophysical Observatory 1.85~m telescope, and 
the Academy of Sciences of the Czech Republic Astronomical Institute 
Ond\v{r}ejov Observatory 2~m telescope. 
We also obtained a smaller set of 22 blue spectra that record 
the variations in the \ion{He}{2} $\lambda 4686$ line that 
probably forms in the focused part of the wind \citep*{gie86b,nin87,kar06b}.
A summary of all these observing runs is given in Table~1 that lists
run number, range of heliocentric Julian dates of observation, 
spectral range recorded, spectral resolving power 
($R=\lambda/\triangle\lambda$, where $\triangle\lambda=$ FWHM
of the line spread function), number of spectra obtained, and 
details about the telescope, spectrograph, and detector. 
The spectra were reduced and transformed to rectified flux 
on a uniform heliocentric wavelength scale (as described in 
\citealt{gie03}). 

\placetable{tab1}      

Most of the red spectra also record the \ion{He}{1}
$\lambda 6678$ absorption line, and we decided to measure the 
stellar radial velocity from this line to check on the orbital 
ephemeris at the time of the {\it HST} observations.  The 
radial velocities were measured in the same way as outlined 
by \citet{gie03} by fitting a Gaussian to the central line 
core.  For the sake of completeness, we also measured the 
radial velocity of the supergiant from the {\it HST} UV 
spectra using a cross-correlation method \citep*{pen99} with
an {\it IUE} spectrum of HD~34078 as the reference template.
The results are presented in Table~2 (given in full in 
the electronic version) that lists the heliocentric Julian 
date of mid-observation, orbital phase, radial velocity and 
its associated error, observed minus calculated velocity 
residual, H$\alpha$ equivalent width, and the corresponding 
run number from Table~1.  We note that independent measurements
of the \ion{He}{1} $\lambda 6678$ line in the Ond\v{r}ejov spectra
using the KOREL package \citep{had07} led to fully consistent
results. 

\placetable{tab2}      

We computed orbital elements from these velocities (omitting those
from {\it HST}) using the non-linear, least-squares fitting 
method of \citet{mor74}.  We made a circular orbital fit with the 
period fixed at the value obtained by \citet{bro99} from data 
spanning a 26 yr interval.  Our results are compared to those from 
\citet{bro99} and from \citet{gie03} in Table~3, and they are 
consistent with these earlier studies.  The current epoch for the 
time of supergiant inferior conjunction $T$(IC) occurs 
$0.017\pm0.012$ days later than the prediction from \citet{bro99}, 
and we will adopt this revised epoch for the definition of 
orbital phase throughout the paper.  We omitted the {\it HST} 
measurements from the orbital solution because of concerns 
about possible systematic differences in the velocities 
derived from the UV lines and \ion{He}{1} $\lambda 6678$, 
but the $(O-C)$ residuals for the {\it HST} measurements 
given in Table~2 show that the UV measurements are in 
reasonable agreement with the velocity curve derived from 
\ion{He}{1} $\lambda 6678$. 

\placetable{tab3}      


\section{Wind and X-ray States During the HST Observations}     

The optical red spectra were obtained with the primary goal of 
monitoring the gas density at the base of the stellar wind of the supergiant. 
The H$\alpha$ observations of HD~226868 are summarized 
in two panels in Figure~1 according to the X-ray state at the time of 
observation (see Fig.~2 below).  The top portions show plots of 
the profiles arranged by orbital phase while the lower grayscale images
show the spectral flux interpolated in radial velocity and orbital 
phase.  The white line in the lower image shows the radial velocity 
curve of the supergiant (Table~3).  These figures show that most of 
the emission/absorption complex appears to follow the orbit of the 
supergiant as expected for an origin in the supergiant wind.  
We measured the H$\alpha$ equivalent width in the same way as before
\citep{gie03} by making a numerical integration over a 40 \AA\ range 
centered on H$\alpha$, and these measurements are listed in 
column 6 of Table~2.  We estimate that the typical measurement error is 
$\pm 0.1$ \AA\ (depending mainly on the S/N ratio of the individual 
spectrum).   

\begin{figure}
\begin{center}
\plottwo{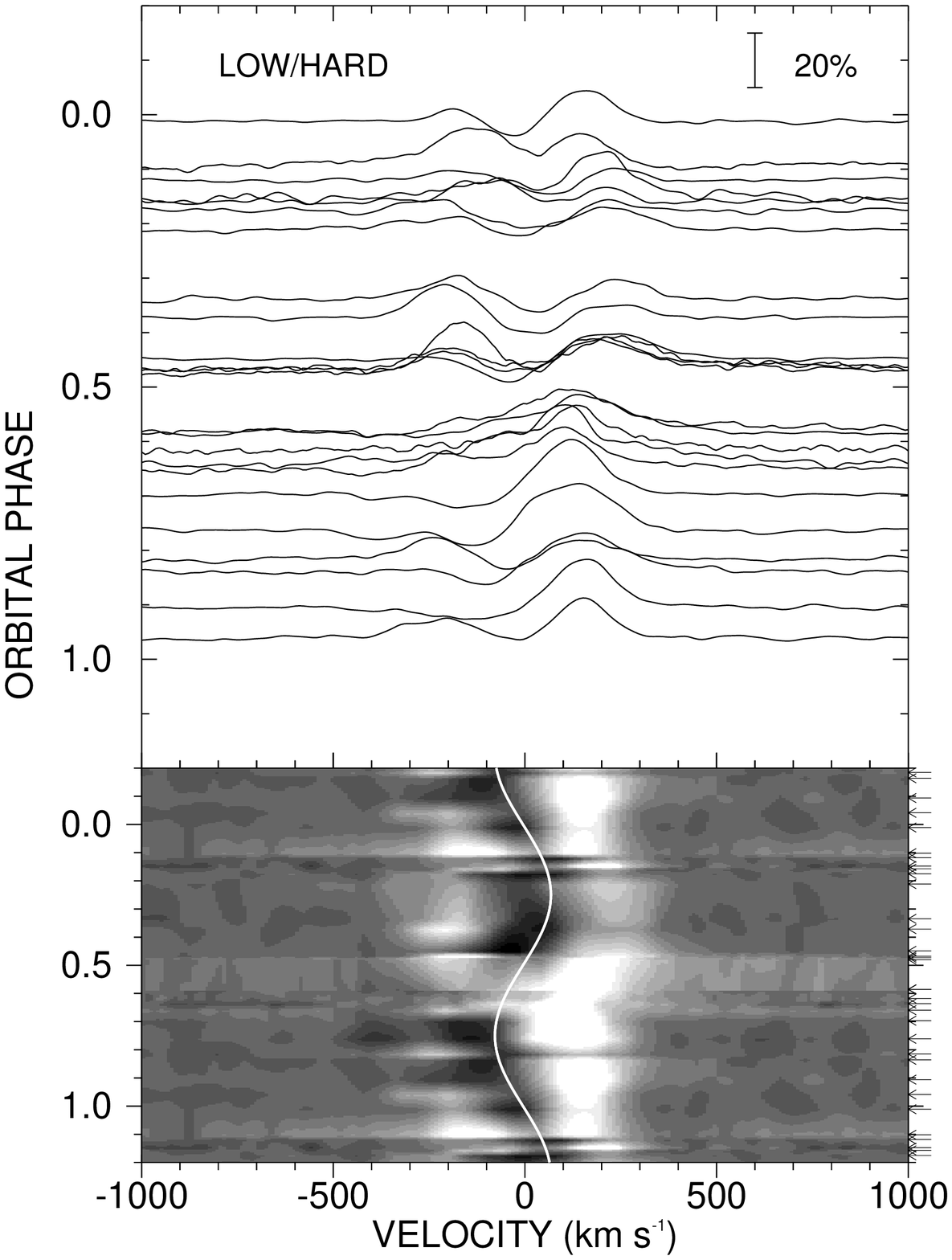}{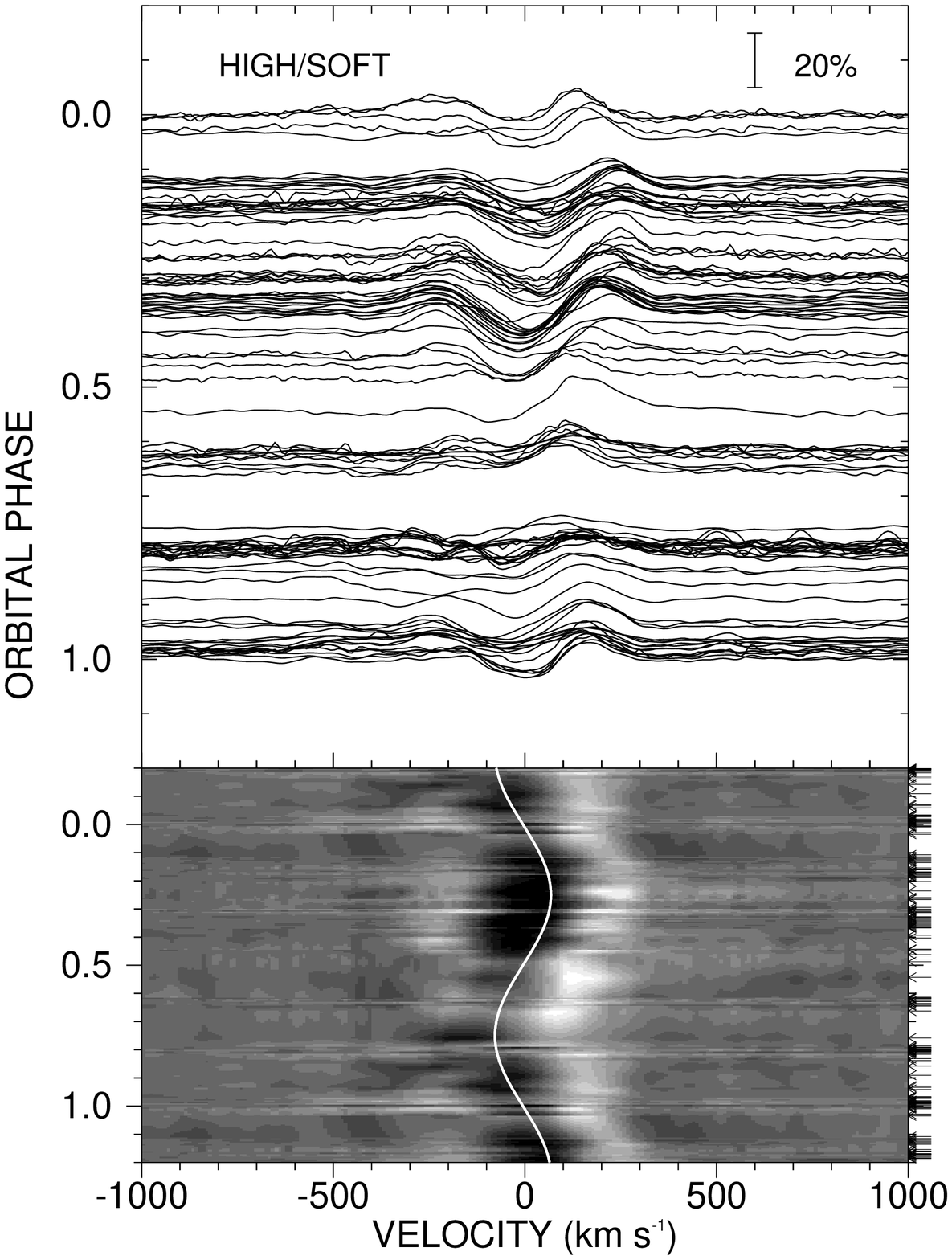}
\end{center}
\caption{Two depictions of the H$\alpha$ profiles as a function 
of heliocentric radial velocity and orbital phase and grouped 
according to X-ray state at the time of observation.  The 
X-ray high/soft group was selected from observations made 
between HJD 2,452,100 -- 2,452,550 and between 
HJD 2,452,770 -- 2,452,880, and all other times were assigned 
to the X-ray low/hard group (see Fig.~2).  The upper 
panel in each shows the profiles with their continua aligned 
to the orbital phase of observation, while the lower panel is 
a grayscale version of the profiles with the first and last $20\%$
of the orbit repeated to improve the sense of phase continuity. 
The grayscale intensities represent the rectified spectral flux between
0.92 (black) and 1.12 (white).  The white line in the grayscale image
shows the orbital velocity curve of the supergiant.}
\label{fig1}
\end{figure}

\placefigure{fig1}     

We show the time evolution of the H$\alpha$ equivalent width for 
a total of 240 measurements from \citet{gie03} and the new observations
in the top panel of Figure~2.  The lower panel of this figure 
shows the daily average soft X-ray flux over the same interval from 
the All-Sky Monitor on the {\it Rossi X-ray Timing Explorer} \citep{lev96}. 
These flux measurements are the quick-look results provided
by the {\it RXTE}/ASM team\footnote{http://xte.mit.edu/}.
The two arrows in the top panel indicate the times of the two 
{\it HST} observing runs, which took place when Cyg~X-1 was 
in the high/soft state.  The new measurements confirm the trends 
described by \citet{gie03} and \citet{tar03} that the H$\alpha$ 
emission tends to be stronger when the soft X-ray flux declines and
that there is a considerable range in emission strength when the 
soft X-ray flux is large (see Fig.~1). 

\begin{figure}
\begin{center}
{\includegraphics[angle=90,height=12cm]{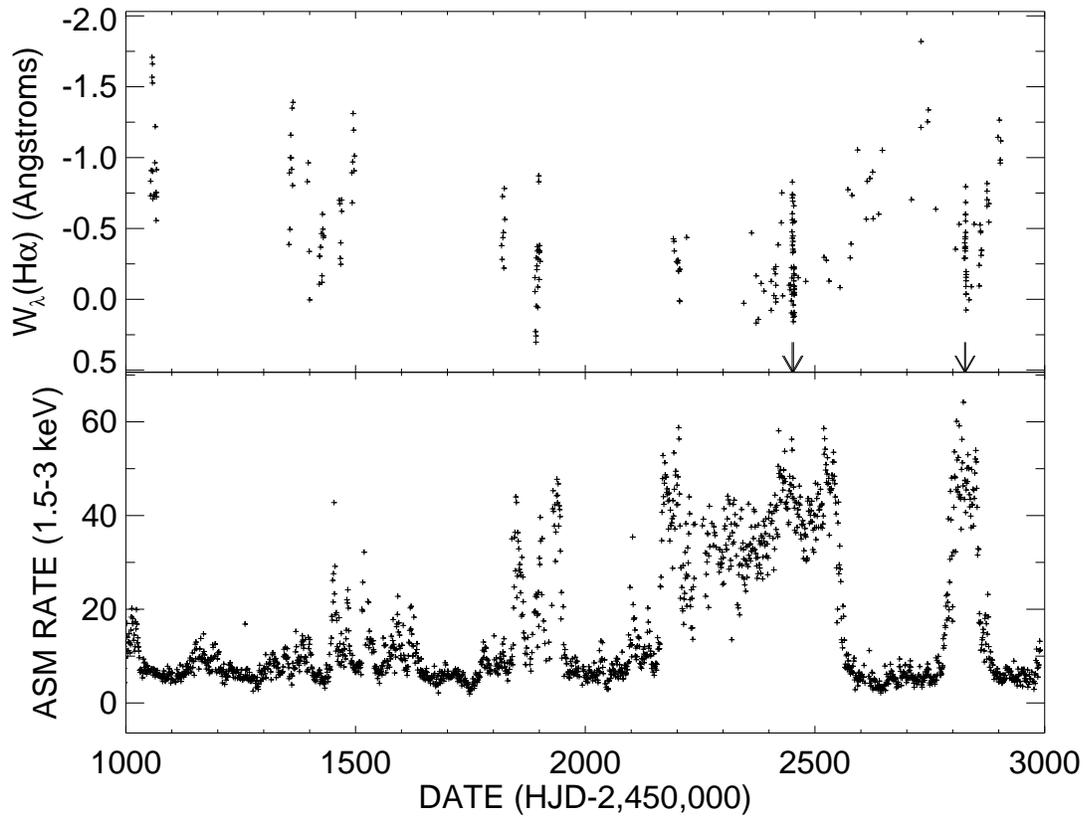}}
\end{center}
\caption{The long term variations in the H$\alpha$ emission strength ({\it above}) 
and daily average soft X-ray flux ({\it below}).  The two arrows in the upper panel 
indicate the times of the STIS observations that occurred during the X-ray 
high/soft state.  The recent, more densely sampled observations show clearly 
how the H$\alpha$ emission increases as the soft X-ray flux declines.}
\label{fig2}
\end{figure}

\placefigure{fig2}     

Figures 3 and 4 show a detailed view of the time evolution of 
the H$\alpha$ emission strength and X-ray flux for the week 
surrounding the {\it HST} runs in 2002 and 2003.  The bottom 
panels in these figures show the X-ray fluxes in both the low 
energy (1.5 -- 3 keV; $+$ signs) and higher energy (5 -- 12 keV;
$\times$ signs) bands for the individual, 90~s exposure, dwell measurements. 
Unfortunately, both the H$\alpha$ and X-ray measurements are 
not exactly coincident in time with the {\it HST} observations, 
so we have made a time interpolation between the closest available 
measurements to estimate the H$\alpha$ emission and X-ray flux 
levels at the times of the {\it HST} observations (summarized 
in Table~4).  All four {\it HST} observations occurred when the 
H$\alpha$ emission was weak and the soft X-ray flux was uniformly strong. 

\begin{figure}
\begin{center}
{\includegraphics[angle=90,height=12cm]{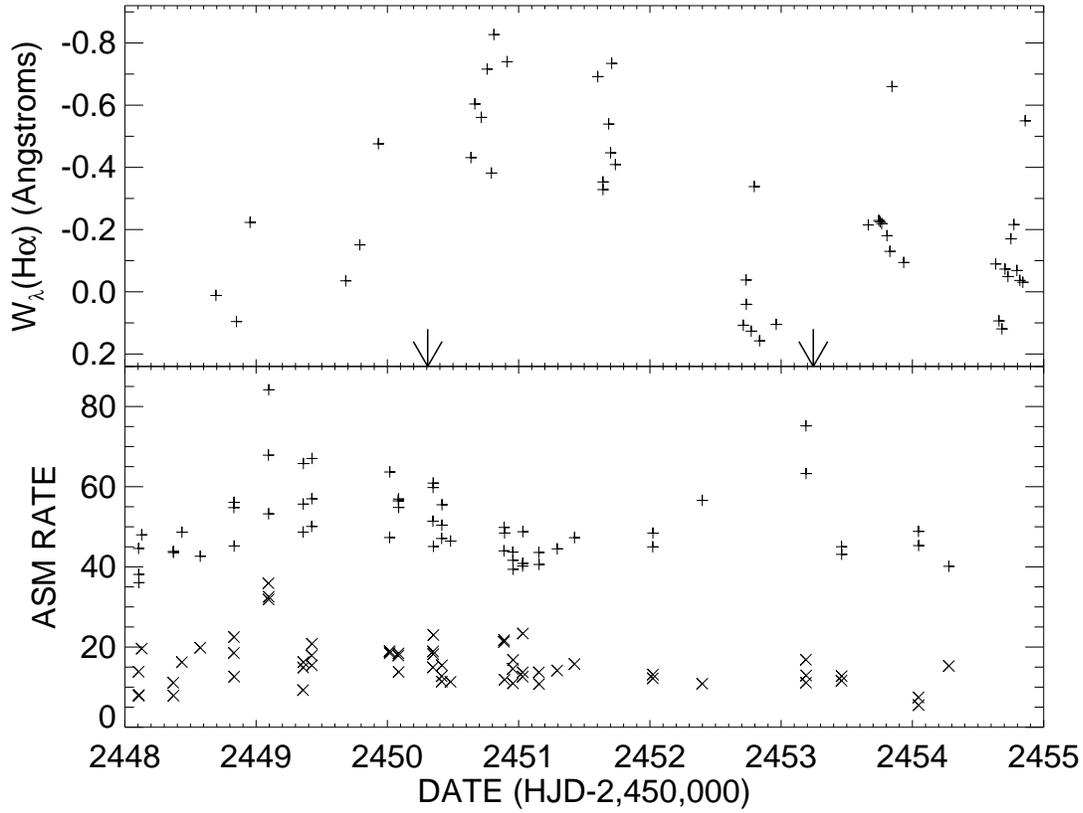}}
\end{center}
\caption{The variations in H$\alpha$ emission strength ({\it above}) and 
X-ray flux for each dwell observation ({\it below}) for the week 
surrounding the first {\it HST} run in 2002.  The arrows in the 
top panel show the times of the STIS observations for each 
orbital conjunction phase.  The symbols in the lower panel indicate 
the count rates in the 1.5 -- 3 keV ($+$) and 5 -- 12 keV ($\times$) bands.} 
\label{fig3}
\end{figure}

\placefigure{fig3}     

\begin{figure}
\begin{center}
{\includegraphics[angle=90,height=12cm]{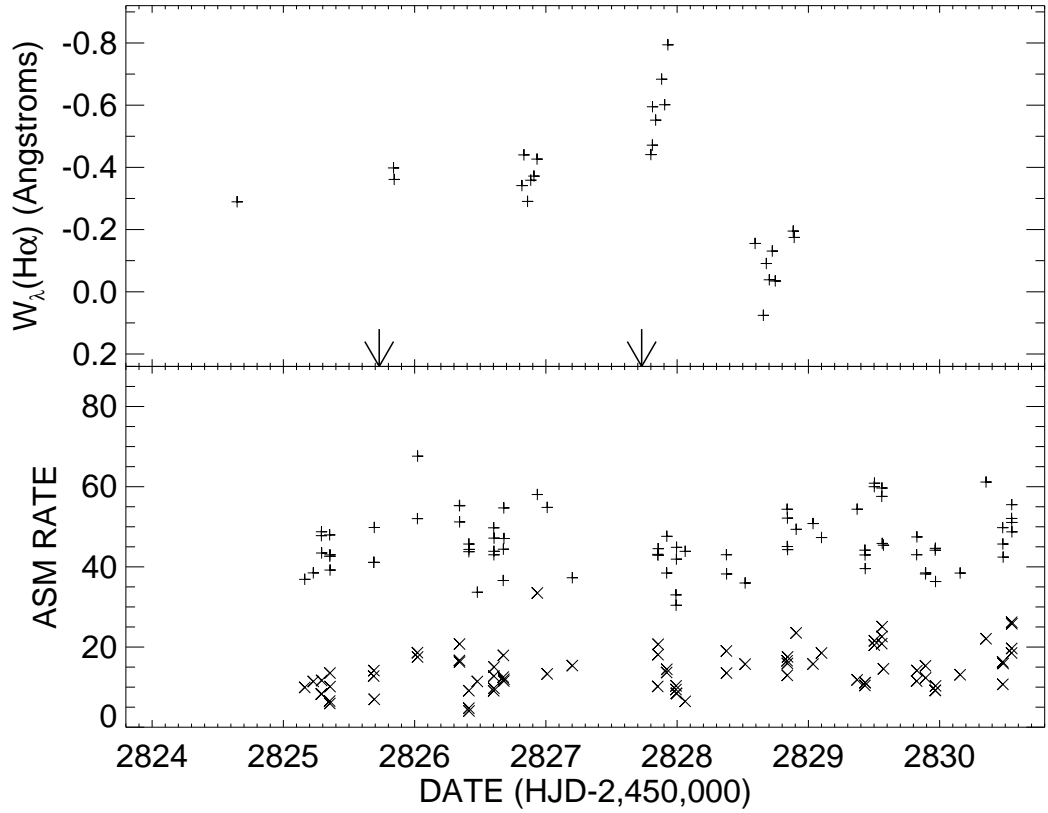}}
\end{center}
\caption{The variations in H$\alpha$ emission strength ({\it above}) and 
X-ray flux for each dwell observation ({\it below}) for the week 
surrounding the second {\it HST} run in 2003 (in the same format 
as Fig.~3).}
\label{fig4}
\end{figure}

\placefigure{fig4}     

\placetable{tab4}      

The other important optical emission line in the spectrum of HD~226868 
is the \ion{He}{2} $\lambda 4686$ feature.  \citet{gie86b} and \citet{nin87}
found that this emission probably forms between the supergiant and 
black hole in a higher density and slower region of the wind, 
the focused wind predicted by \citet{fri82}.  We obtained a limited 
number of spectra of the \ion{He}{2} $\lambda 4686$ feature (runs 8 and 9
in Table~1) at the times of the two {\it HST} observation sets, and these
observations show profile variations with orbital phase that are quite 
similar to those seen previously \citep{gie86b,nin87,kar06b}.  
We present in Table~5 the observed equivalent width of the 
\ion{He}{2} $\lambda 4686$ emission/absorption complex 
(made by numerical integration between 4677 and 4695 \AA )
for spectra from these two runs and from a third run 
(\#10 in Table~1) that occurred when the 
system had returned to the X-ray low/hard state.  There was no 
significant difference in the amount of \ion{He}{2} $\lambda 4686$ 
emission present between the times of the two X-ray states.  
We compare in Table~6 the averages of these measurements with earlier 
orbital phase averages \citep{gie86a,bro99} and time averages 
\citep{nin87} of the emission strength that correspond mainly to 
times when the system was in the low/hard X-ray state.  
Note that \citet{nin87} report only the net emission equivalent 
width after subtraction of the photospheric absorption profile 
of the similar star HD~149038, and we estimate that this 
procedure increased the emission strength by $0.25$\AA ~according 
to the plot of the spectrum of HD~149038 given in Figure~1 of 
\citet{nin87}.   We find that the \ion{He}{2} $\lambda 4686$
emission has remained more or less constant in strength
over the decades of observation and between the X-ray states, 
and this suggests the tidal gas stream towards the black hole 
experiences much less variation than does the global wind 
as observed in the H$\alpha$ emission line. 

\placetable{tab5}      

\placetable{tab6}      


\section{Orbital Variations in the UV Wind Lines}     

Our primary interest here is how the X-ray flux ionizes  
portions of the supergiant's wind and how our line of sight 
through the ionized zones changes with orbital phase. 
We show in Figures 5, 6, and 7 the changes observed between 
conjunctions in the major UV wind lines of 
\ion{N}{5} $\lambda\lambda 1238, 1242$,
\ion{Si}{4} $\lambda\lambda 1393, 1402$, and 
\ion{C}{4} $\lambda\lambda 1548, 1550$.  
In a companion paper, \citet{vrt08} show the variations 
observed in several other weaker lines.  
The top panel in each of these figures shows the variations in 2002 
between the phases with the black hole behind ($\phi = 0.0$; {\it solid line})
and in the foreground of the supergiant ($\phi = 0.5$; {\it dotted line}), and 
the bottom panel shows the same for the 2003 run.  The spectra are 
plotted as a function of radial velocity for the shorter wavelength
component in the frame of the supergiant (according to the orbital 
solution in Table~3).  All three of these transitions display a 
large reduction in the extent and depth of the blueshifted 
absorption component when the black hole is in the foreground 
(the Hatchett-McCray effect).  These changes reflect the X-ray photoionization
and resulting superionization of these ions in the wind gas seen projected 
against the supergiant.  We also find some evidence of the associated reduction
in the strength of the red emission component due to the loss of these 
gas ions in the outflow away from our line of sight when the black hole is 
in the background.  The variations between conjunctions appear to 
be almost identical for the observations in 2002 and 2003, which is 
probably due to the very similar X-ray fluxes that existed at those
times of observation (\S3, Table~4). 

\begin{figure}
\begin{center}
{\includegraphics[angle=90,height=12cm]{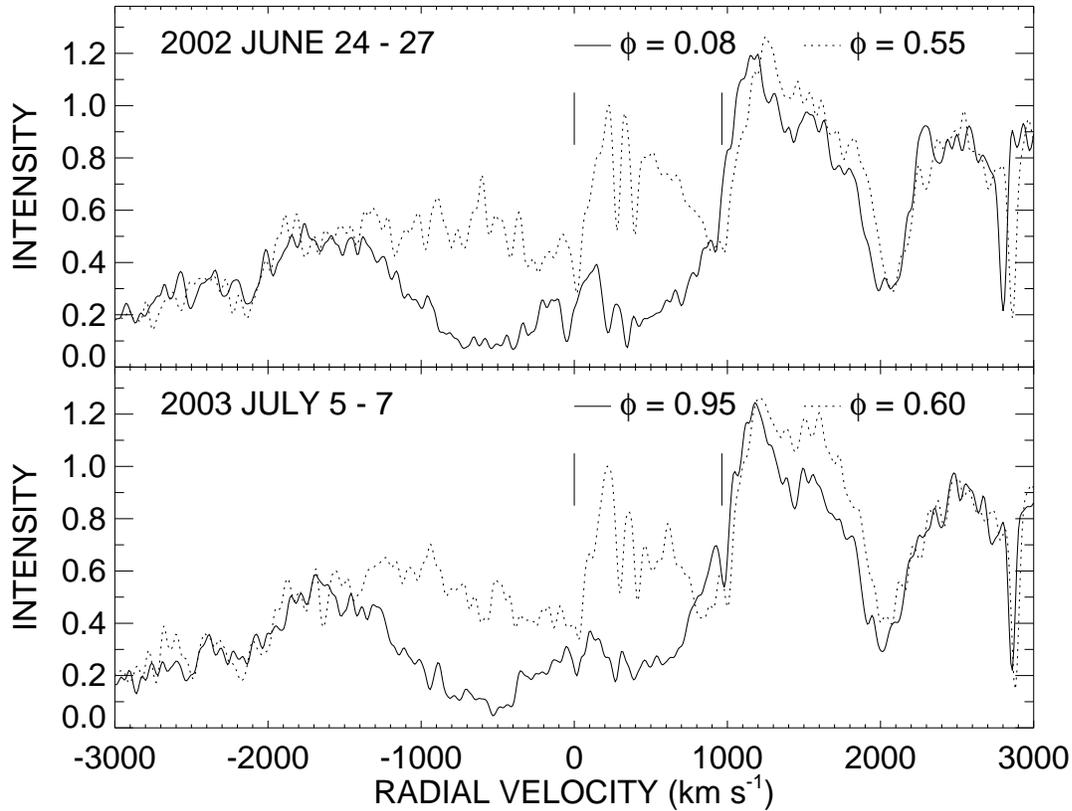}}
\end{center}
\caption{The spectral variations observed between conjunctions
for the \ion{N}{5} $\lambda\lambda 1238, 1242$ wind feature
in both the 2002 ({\it top}) and 2003 ({\it bottom}) runs.  The spectra 
are plotted as a function of Doppler shift for the blue component
of the doublet in the rest frame of the supergiant (which leads to 
small offsets in the positions of the narrow, interstellar lines that
are stationary in the absolute frame).  Vertical line segments indicate 
the rest wavelength positions of both components of the doublet.}   
\label{fig5}
\end{figure}

\placefigure{fig5}     

\begin{figure}
\begin{center}
{\includegraphics[angle=90,height=12cm]{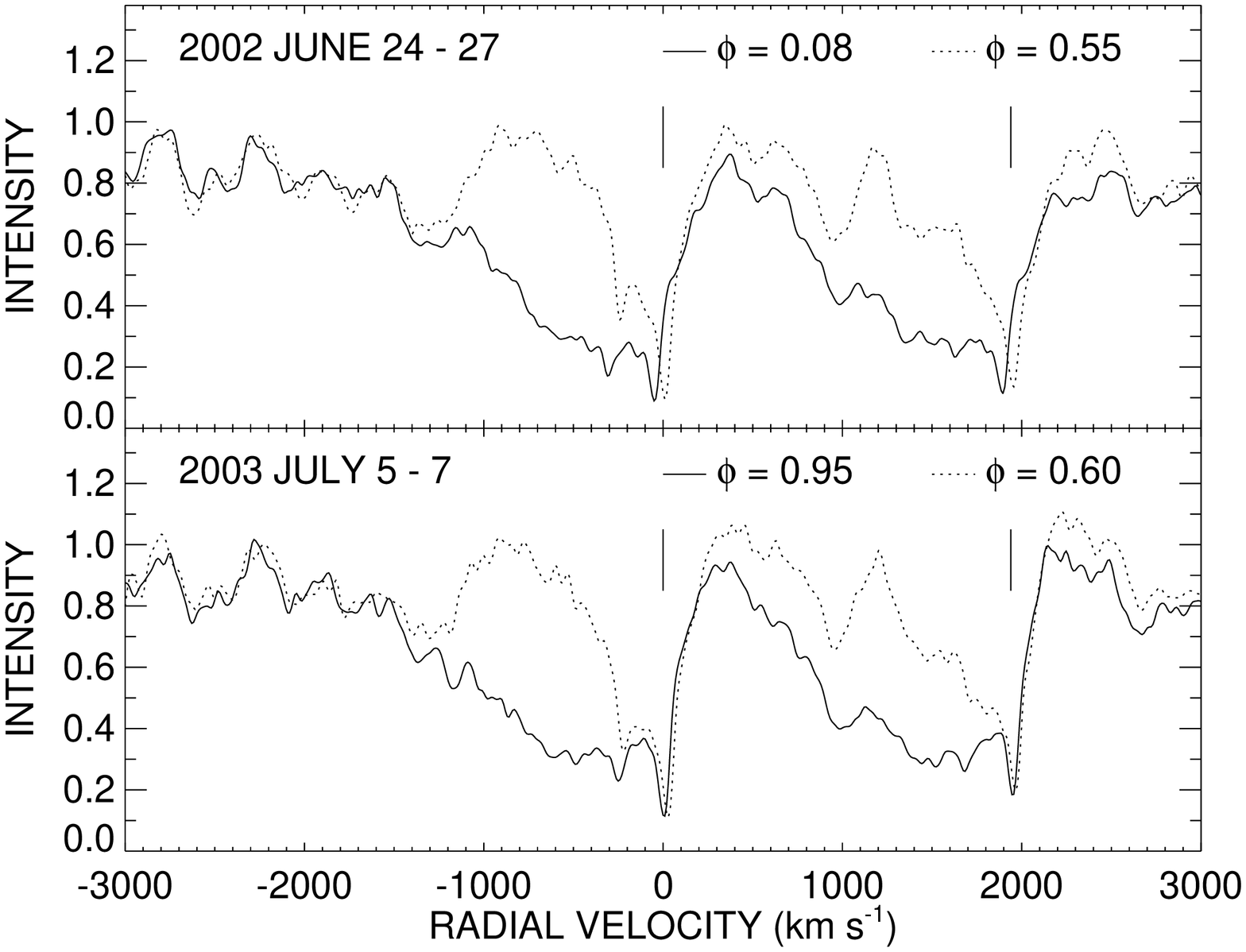}}
\end{center}
\caption{The spectral variations observed between conjunctions
for the \ion{Si}{4} $\lambda\lambda 1393, 1402$ wind feature
in the same format as Fig.~5.}
\label{fig6}
\end{figure}

\placefigure{fig6}     

\begin{figure}
\begin{center}
{\includegraphics[angle=90,height=12cm]{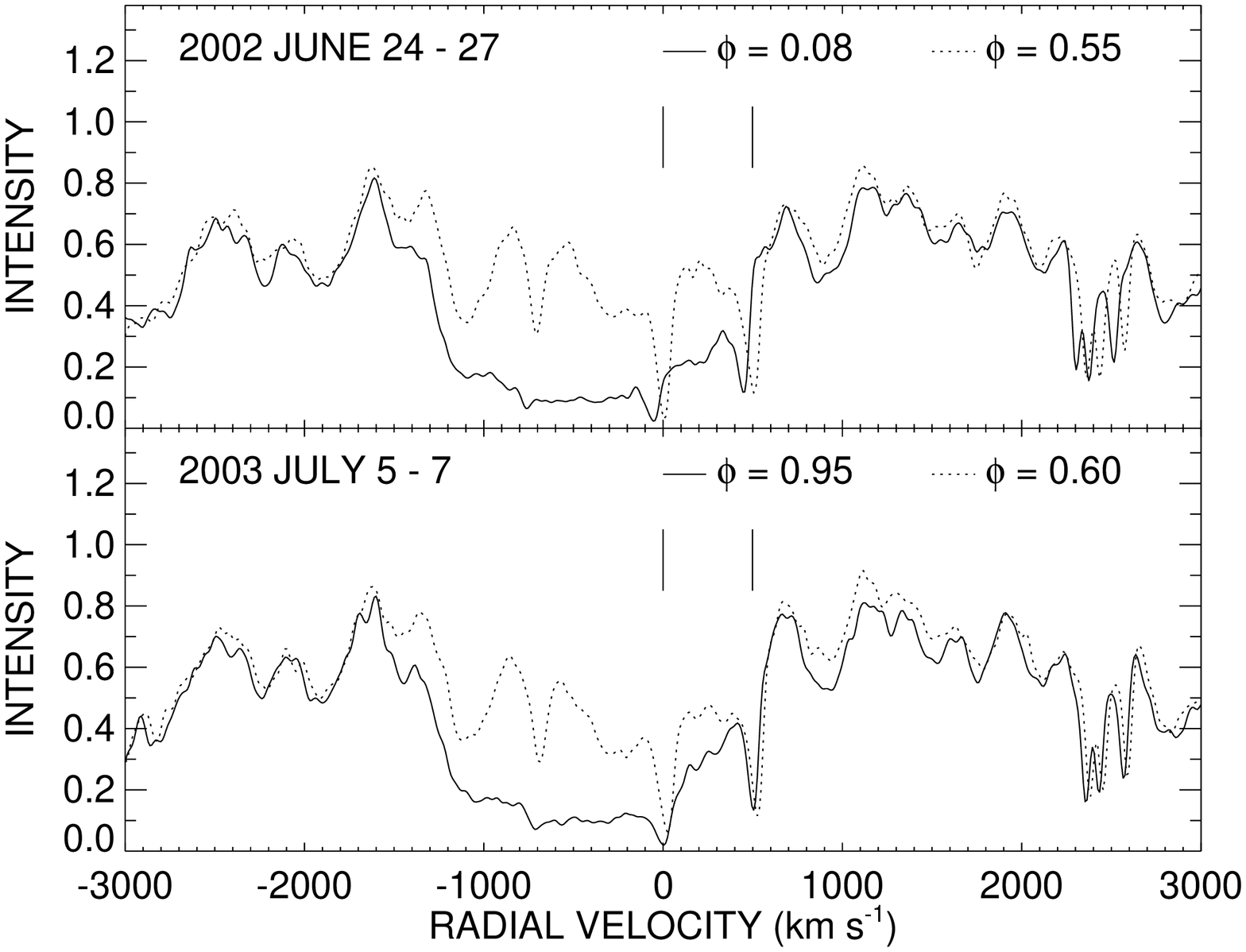}}
\end{center}
\caption{The spectral variations observed between conjunctions
for the \ion{C}{4} $\lambda\lambda 1548, 1550$ wind feature
in the same format as Fig.~5.}
\label{fig7}
\end{figure}

\placefigure{fig7}     

The shapes of the wind profiles near orbital phase $\phi = 0.5$ 
suggest that the P~Cygni absorption troughs have almost entirely 
disappeared, or that the wind ionization extends all the way 
towards the exposed photosphere of the supergiant.  We made some 
simple calculations of the appearance of wind profiles that arise only 
from the gas hidden from the X-ray source, the so-called 
shadow wind \citep{blo94} that is illustrated diagrammatically in Figure~8. 
Here we assume that the normal supergiant wind is confined to the 
region where the line of sight to the black hole is blocked by the 
supergiant.  However, we expect that in reality this shadow region
is partially exposed to X-rays by wind scattering (probably 
comparable to the scattering along our line of sight through 
the wind, i.e., a few percent of the unobscured X-ray flux; 
\citealt{wen99}) and by the Coriolis deflection that will bring
the shadowed gas into regions of X-ray illumination \citep{blo94}.
The line synthesis calculations are based on a modification of
the Sobolev-Exact Integration method \citep*{lam87} that was 
developed by \citet{vlo01}.  The simplifying assumption in this 
model is that the observer lies along the axis joining the 
stars at the two conjunctions (or that the orbital inclination 
is $i=90^\circ$ and the orbital phases are $\phi=0.0,0.5$).   
However, since the actual inclination is smaller 
($i = 33^\circ - 40^\circ$; \citealt*{gie86a,bro02}), 
our line of sight at the conjunctions will 
include somewhat different portions of the occulted and unocculted 
wind (Fig.~8), so these models are first approximations of the predicted 
variations for a shadow wind.  In a companion paper \citep{vrt08}, 
we present a more complete calculation based upon a realistic 
orbital inclination and the method outlined by \citet{bor99}. 

\begin{figure}
\begin{center}
{\includegraphics[angle=90,height=12cm]{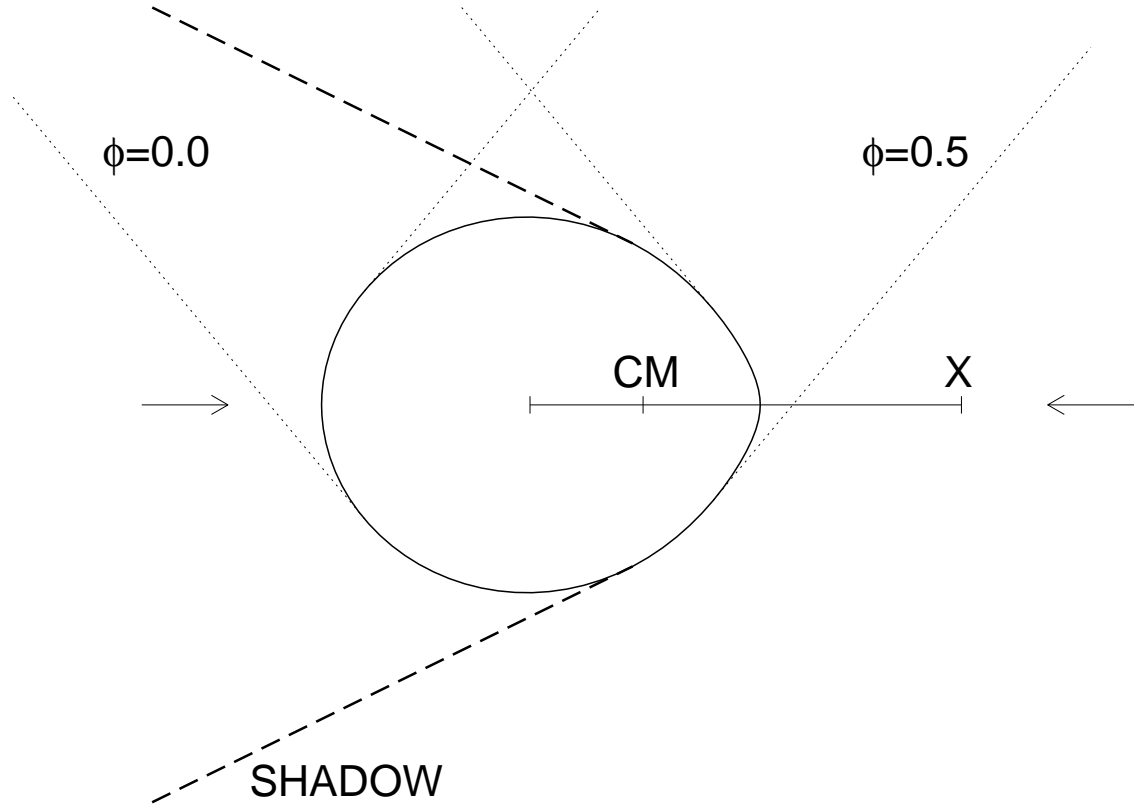}}
\end{center}
\caption{A cartoon view of the binary from within the orbital plane.
The normal wind ions are assumed to exist only within the 
shadow wind zone to the left of the supergiant 
(bounded by a thick, dashed line) where the X-ray 
flux from the vicinity of the black hole (marked by ``X'') is fully blocked, 
while the gas is ionized to higher levels in the rest of the wind. 
The simple model predictions are based upon viewing the system 
along the line of centers (shown by arrows), while in fact we 
observe the supergiant at a lower inclination angle 
($i=40^\circ$ assumed).  The dotted lines show the part of
the line of sight that is projected onto the disk of the
star at the two orbital phases for an observer at the correct
direction with respect to the orbital plane. 
The shadow wind region projected against the star in each case 
is the area bounded by the surface of the star and the dashed 
and dotted line for that phase. 
CM marks the center of mass position.}
\label{fig8}
\end{figure}

\placefigure{fig8}     

The line synthesis is based upon a set of adopted parameters 
and two fitting parameters.  Most of the adopted parameters 
come from the study of Cyg~X-1 and similar X-ray binaries by 
\citet{vlo01}, and in particular, we assume a wind velocity 
law exponent of $\gamma=1$ (eq.~2 in \citealt{vlo01}),
a semimajor axis equivalent to $2.1 R_\star$ (where 
$R_\star$ is the radius of the supergiant; \citealt{gie86a}),
and a characteristic turbulent velocity in the wind equal to 
$0.1 v_\infty$ \citep*{gro89}
where $v_\infty$ is the terminal velocity in the 
undisturbed wind.  The photospheric components corresponding to the wind
transitions were assumed to be Gaussian in shape with parameters 
set by Gaussian fits of the photospheric profiles in the 
non-LTE, line blanketed model spectra of \citet{lan03}
(for $T_{\rm eff}=30000$~K, $\log g = 3.0$, $V\sin i = 100$ km~s$^{-1}$, 
a linear limb darkening coefficient of $\epsilon=0.50$, and a
spectral resolving power of 10000).  

The final two parameters, the wind terminal velocity $v_\infty$ and
the integrated optical depth through the shadow wind of the blue component
of the transition $\tau$, were fit by trial and error in order to 
match the entire set of the three wind lines at each conjunction. 
Note that the value of $v_\infty$ is set mainly by fits of 
the profiles at orbital phase $\phi=0.0$ when the undisturbed wind 
is in the foreground and projected against the star. 
The best match was made with $v_\infty = 1200$ km~s$^{-1}$ and 
$\tau = 6$, 4, and 10 for the \ion{N}{5}, \ion{Si}{4}, \ion{C}{4}
features, respectively.  The shadow wind model profiles are presented 
in Figures 9, 10, and 11 for these three wind lines.  The top panel 
in these figures shows the model ({\it diamonds}) and average observed 
spectra ({\it solid line}) for $\phi=0.0$ while the lower panel 
illustrates the same for phase $\phi=0.5$ (black hole in the foreground). 
The model profiles were renormalized in flux in each case to 
match the pseudo-continuum beyond the line wings in the observed spectra. 
Each panel also contains the predicted photospheric spectrum for 
the region from the models of \citet{lan03}. 
These representative fits successfully reproduce many of the
profile characteristics, especially when line blending from 
the other photospheric lines is taken into account.  This 
suggests that the geometry of the photoionization zone is probably 
not too different from that assumed for a shadow wind (at least 
for the high/soft X-ray state).  However, we caution again that 
our fitting parameters are based upon an axial viewing orientation
while our actual view is more oblique.  For example, according to the 
geometry sketched in Figure~8 (for $i=40^\circ$), the fastest moving 
part of the shadow wind that is projected against the supergiant 
at $\phi=0.0$ occurs at a radial distance from 
the center of the supergiant of $\approx 4.6 R_\star$ where the wind
has not yet reached terminal velocity.  Thus, our fit value of 
$v_\infty = 1200$ km~s$^{-1}$ is probably well below the actual value 
(which is probably closer to 1600 km~s$^{-1}$).

\begin{figure}
\begin{center}
{\includegraphics[angle=90,height=12cm]{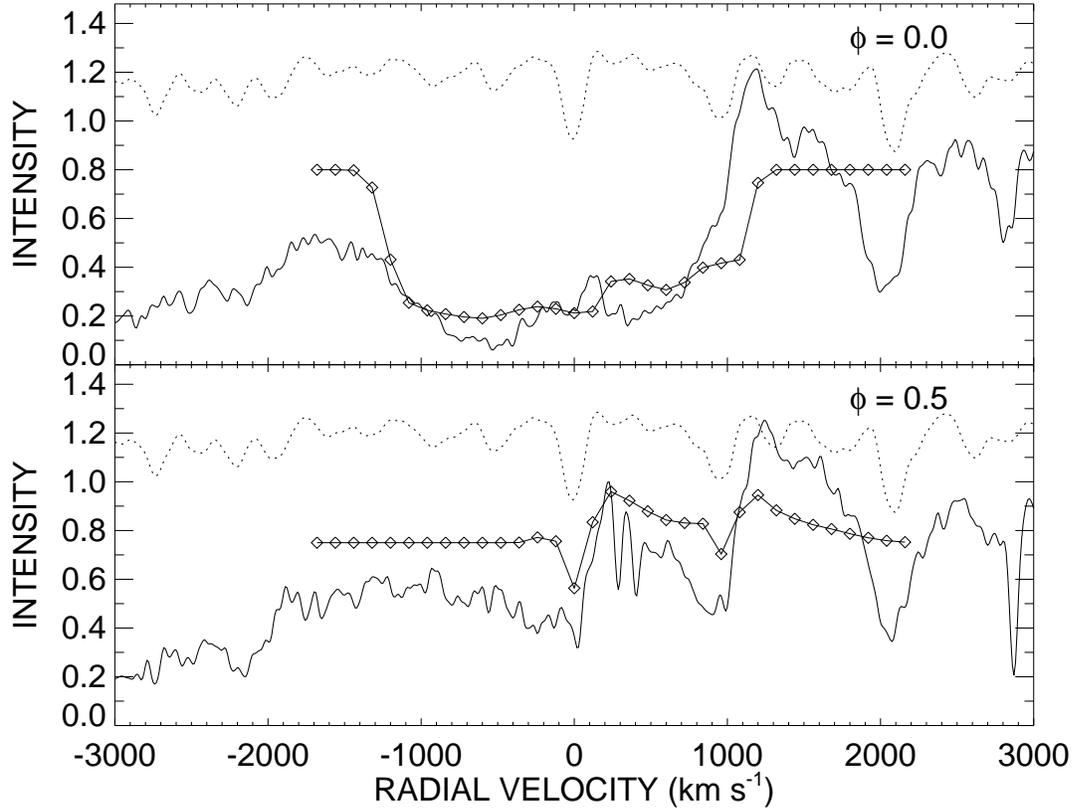}}
\end{center}
\caption{The average observed spectrum ({\it solid line}) and the 
shadow wind model spectrum ({\it connected diamonds}) for the 
\ion{N}{5} $\lambda\lambda 1238, 1242$ wind feature  
(plotted as a function of Doppler shift for the blue component
of the doublet in the rest frame of the supergiant).   The top panel
shows the profiles for case where the black hole is in the background
while the bottom profile illustrates the other conjunction case 
where the black hole is in the foreground (photoionizing the 
wind gas seen projected against the supergiant).  The dotted line 
shows the predicted photospheric line spectrum (offset for clarity)
from the models of \citet{lan03}.  Note that the latter representation 
does not include the broad Ly$\alpha$ wings that depress the continuum 
in the observed spectrum towards the left hand side of the diagram.} 
\label{fig9}
\end{figure}

\placefigure{fig9}     

\begin{figure}
\begin{center}
{\includegraphics[angle=90,height=12cm]{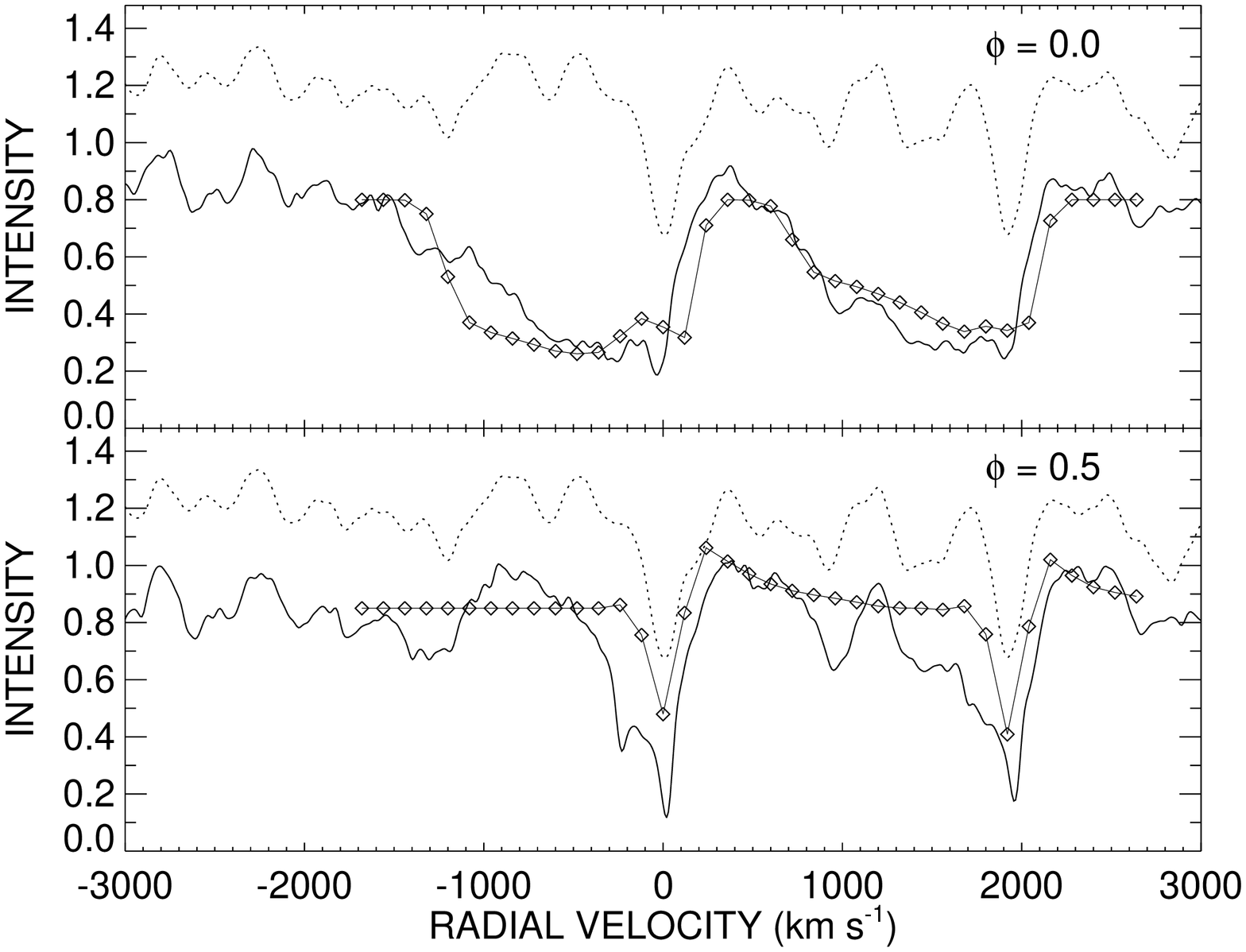}}
\end{center}
\caption{The average observed spectrum ({\it solid line}) and the 
shadow wind model spectrum ({\it connected diamonds}) for the 
\ion{Si}{4} $\lambda\lambda 1393, 1402$ wind feature in the same format
as Fig.~9.}
\label{fig10}
\end{figure}

\placefigure{fig10}    

\begin{figure}
\begin{center}
{\includegraphics[angle=90,height=12cm]{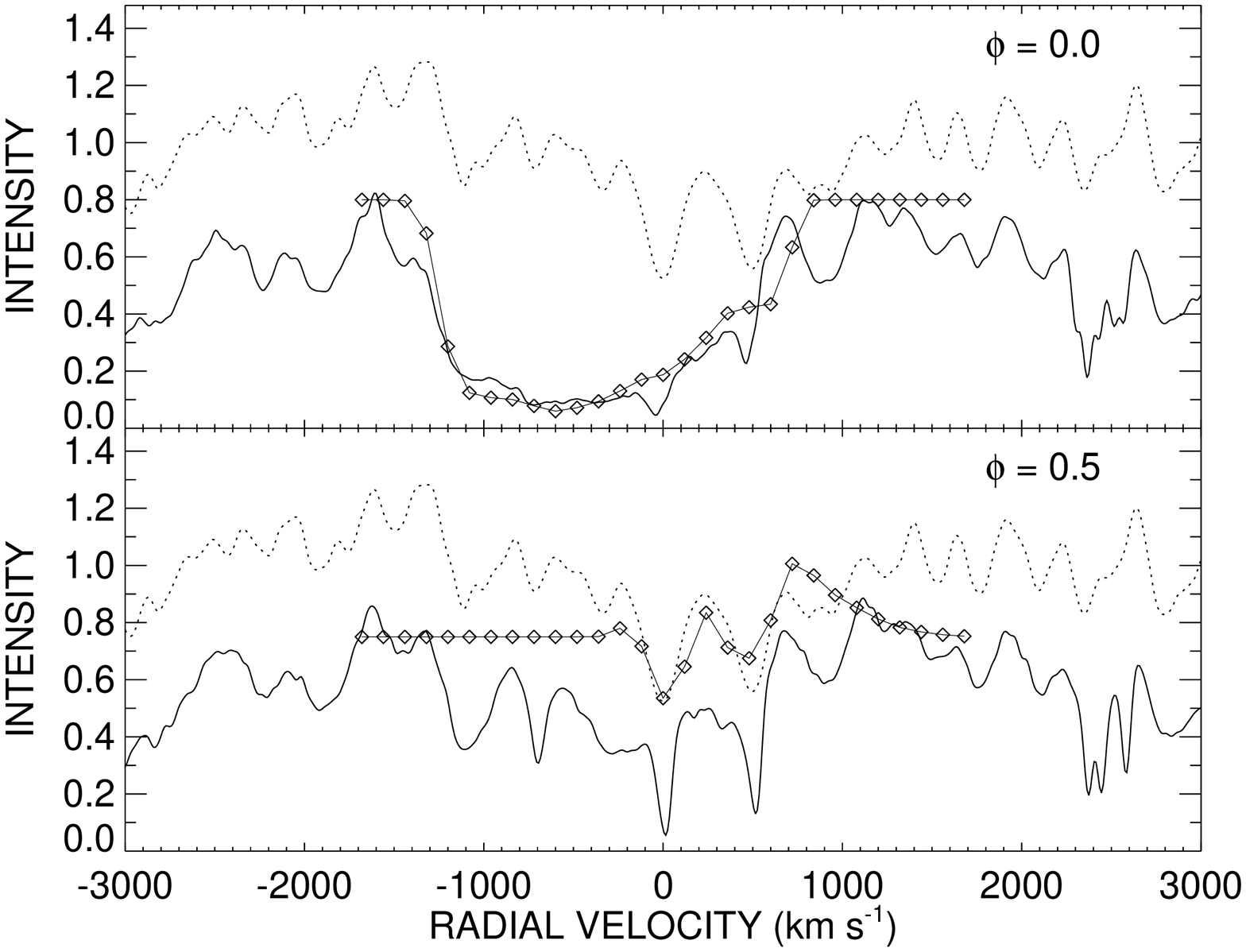}}
\end{center}
\caption{The average observed spectrum ({\it solid line}) and the 
shadow wind model spectrum ({\it connected diamonds}) for the 
\ion{C}{4} $\lambda\lambda 1548, 1550$ wind feature in the same format
as Fig.~9.}
\label{fig11}
\end{figure}

\placefigure{fig11}    

The low orbital inclination and the subsequent limited projection of the 
shadow wind region against the disk of the photosphere results in 
P~Cygni absorption troughs that are unusually weak for the spectra of 
supergiants like HD~226868.  We show in Figures~12 and 13 montages of
the \ion{Si}{4} and \ion{C}{4} wind features in four other O9.7~Iab 
supergiants as seen in high dispersion spectra from the {\it IUE} archive.
The mean spectrum of HD~226868 at orbital phase $\phi=0.0$ (shown 
at the top of both figures) shows that the wind profiles have a lower
optical depth and attain a smaller blue-shifted velocity because 
only a small portion of the shadow wind is projected against the 
disk of the supergiant at its inferior conjunction (see Fig.~8).  

\begin{figure}
\begin{center}
{\includegraphics[angle=90,height=12cm]{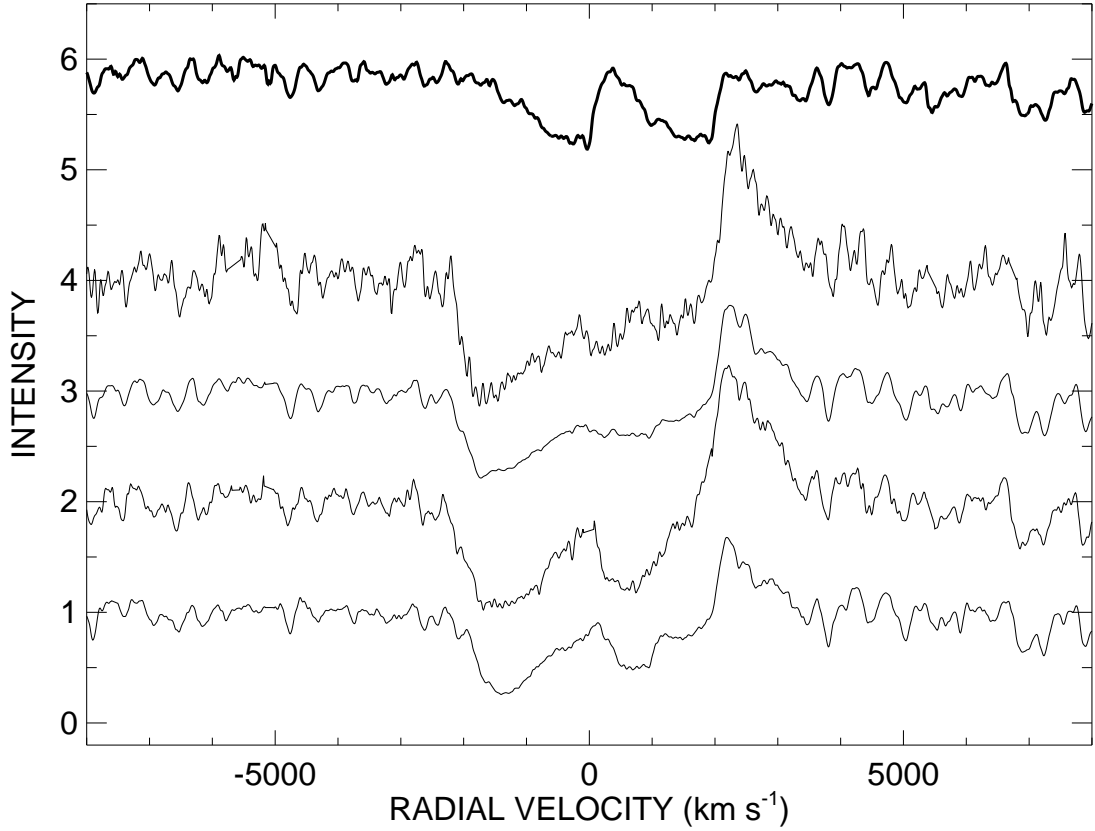}}
\end{center}
\caption{The mean \ion{Si}{4} wind profile for orbital phase $\phi=0.0$
as a function of radial velocity for the blue member of the doublet.  
The spectrum of HD~226868 is plotted at the top as a thick line,
and below are average spectra for other O9.7~Iab supergiants
(offset in intensity for clarity).  From top to bottom, these are
HD~75222, HD~152003, HD~149038, and HD~167264.}
\label{fig12}
\end{figure}

\placefigure{fig12}    

\begin{figure}
\begin{center}
{\includegraphics[angle=90,height=12cm]{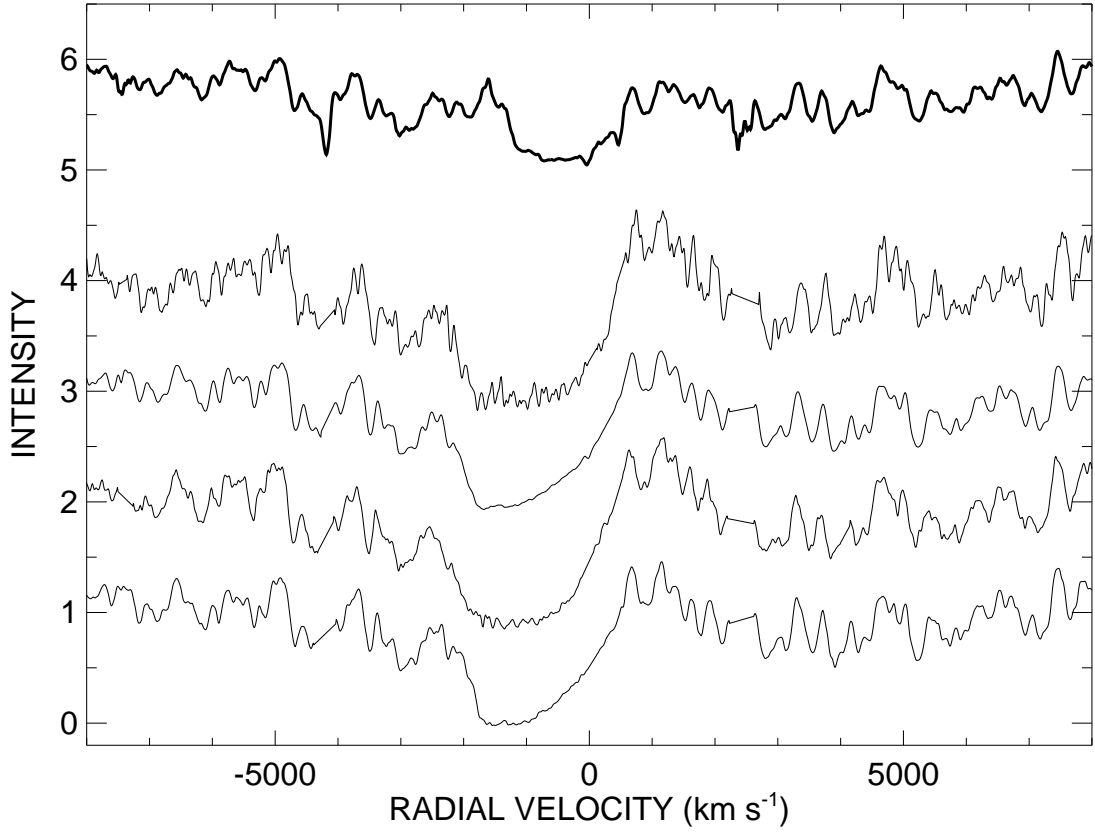}}
\end{center}
\caption{The mean \ion{C}{4} wind profile of HD~226868 for orbital phase 
$\phi=0.0$ ({\it top, thick line}) compared to those of other O9.7~Iab 
supergiants (in the same format as Fig.~12).}
\label{fig13}
\end{figure}

\placefigure{fig13}    


\section{Comparison of the Wind Lines in the X-ray Low and High States}     

Our original goal was to obtain another set of STIS spectra when 
Cyg~X-1 was in the low/hard X-ray state in order to determine how
the wind ionization conditions change with X-ray state.  With the loss of STIS, 
such a comparison is not possible at present.  However, the low 
dispersion FUV spectra of HD~226868 made with {\it IUE} were in 
most cases made when Cyg~X-1 was in the low/hard state.  Thus, 
we can investigate differences in the wind ionization properties 
between X-ray states through a comparison of the {\it HST} 
high/soft state spectra with the {\it IUE} low/hard state spectra. 

We collected 30, low dispersion, short wavelength prime camera 
spectra of HD~226868 from the {\it IUE} archive, and transformed
these to rectified flux versions on a uniform wavelength grid. 
Next, we smoothed, rectified, and rebinned the {\it HST} STIS spectra 
onto the same {\it IUE} wavelength grid so that the line blended 
structures would appear the same as they do in the {\it IUE} 
spectra.  We then measured the effective absorption strength 
by determining the mean flux across a spectral range that 
extends over the full range of the apparent wind feature 
(as done by \citealt{vlo01}).  This average flux will reflect 
both the changing P~Cygni absorption and the other line blends 
(including interstellar components), but since the latter are 
generally constant in time, the average flux will serve to show the 
relative variations in the wind absorption (low flux when the 
P~Cygni trough is deep and high flux when the trough weakens). 

The average flux measurements for both the {\it IUE} and 
rebinned {\it HST} spectra are given in Table~7, which lists 
the heliocentric Julian date of mid-exposure, orbital phase 
(from Table~3), the mean rectified flux across the 
\ion{Si}{4} and \ion{C}{4} wind lines, the telescope of origin, 
the X-ray state at the time of the observations, and a 
code for references discussing the contemporary X-ray fluxes.  
Note that we did not measure the \ion{N}{5} transition 
in the {\it IUE} spectra because these spectra are poorly 
exposed at the short wavelength end.  The {\it IUE} average flux 
measurements have a typical error of $\pm 7\%$ based upon 
the scatter in the results from closely separated pairs of spectra. 

\placetable{tab5}      

The average fluxes across the wind lines are plotted as a function 
of orbital phase in Figures 14 and 15 for \ion{Si}{4} and \ion{C}{4}, 
respectively (see a similar depiction in Fig.~1 from \citealt{vlo01}). 
Different symbols show these measurements for the different 
X-ray states as observed with {\it IUE} and {\it HST}. 
We suspect that despite our efforts to rectify the {\it IUE} and {\it HST}
spectra in the same way, there are probably still some systematic 
differences since the mean fluxes for the {\it HST} spectra appear
to be somewhat lower than those for the {\it IUE} spectra.  
Nevertheless, the amplitude of line strength variation appears to be 
more or less the same in each of the {\it IUE} low/hard state, 
{\it IUE} high/soft state, and {\it HST} high/soft state spectra. 
This result indicates that the Hatchett-McCray effect (and the amount of 
wind photoionization it represents) occurs at about the same level in 
both X-ray states. 

\begin{figure}
\begin{center}
{\includegraphics[angle=90,height=12cm]{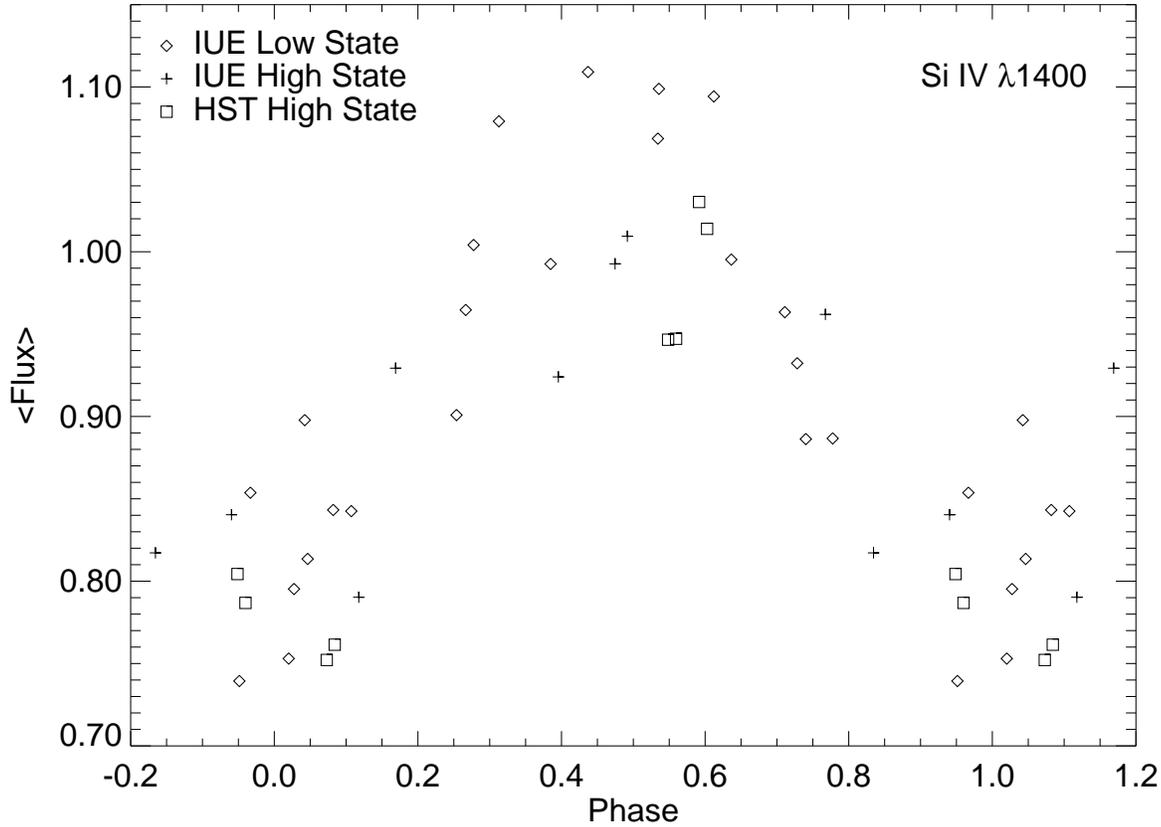}}
\end{center}
\caption{The mean flux across the \ion{Si}{4} wind profile plotted
as a function of orbital phase.  The different symbols represent 
measurements of {\it IUE} low/hard state ({\it diamonds}), 
{\it IUE} high/soft state ({\it plus signs}), and 
{\it HST} high/soft state spectra ({\it squares}). The Hatchett-McCray
effect is seen as low flux (deep absorption) when the black hole is 
in the background and high flux (little absorption) when the black hole 
is in the foreground.  The {\it IUE} flux measurements have a typical
error of $\pm 7\%$.}
\label{fig14}
\end{figure}

\placefigure{fig14}    

\begin{figure}
\begin{center}
{\includegraphics[angle=90,height=12cm]{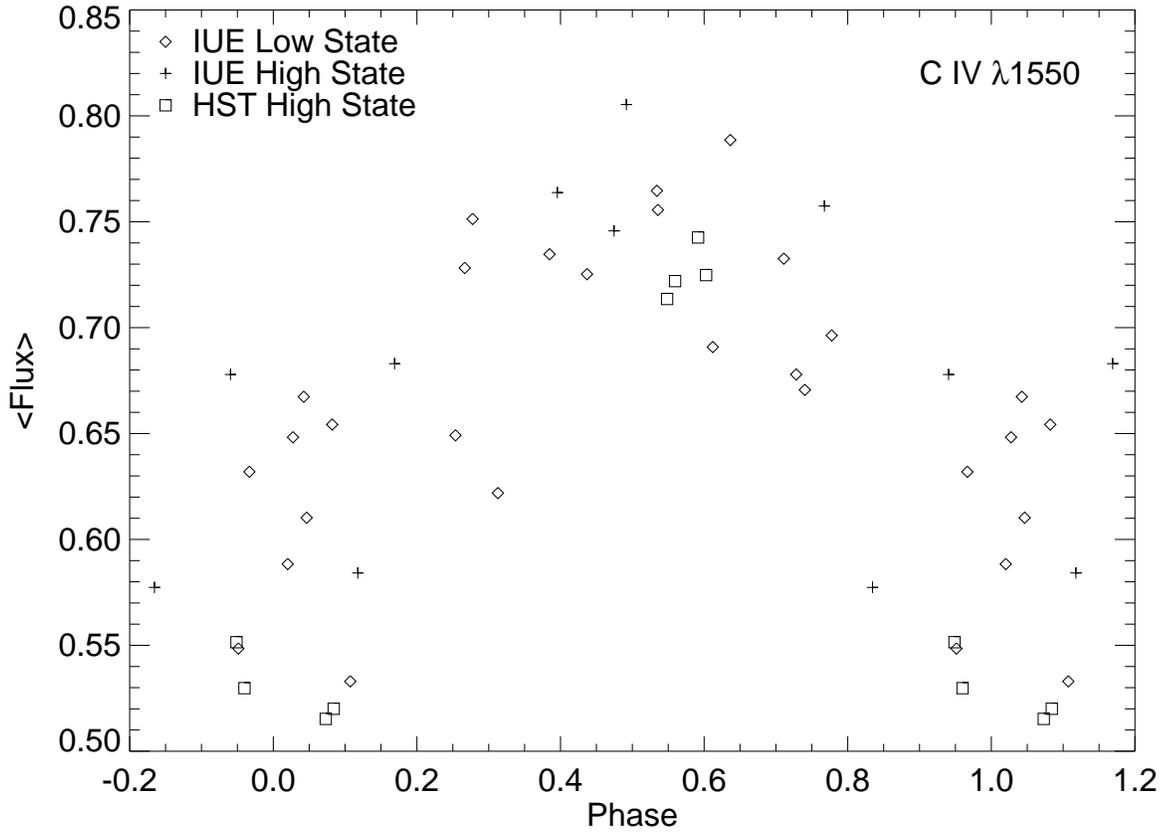}}
\end{center}
\caption{The mean flux across the \ion{C}{4} wind profile plotted
as a function of orbital phase in the same format as Fig.~14.}
\label{fig15}
\end{figure}

\placefigure{fig15}    


\section{Discussion}     

The X-ray accretion flux of Cyg X-1 is fueled by mass transfer
from the supergiant.  We argue in this section that the mass 
transfer process is dominated by a wind focused along the axis 
joining the stars.  However, the accretion of this gas by 
the black hole may be influenced by the strength of 
the radiatively driven shadow wind that is directed away 
from the black hole.  We begin by reviewing the 
most pertinent observational results from this investigation, 
and then we consider the interplay between the dynamics of the 
wind outflow and the X-ray accretion flux. 

First, the {\it HST} STIS spectra of HD~226868 that we obtained at 
two epochs when the system was in the high/soft state show dramatic 
variations in the wind line strength that result from a 
superionization of the gas atoms illuminated by the X-ray flux. 
Shadow wind models, in which the wind ions only exist in the region 
where X-rays are blocked by the supergiant, make a reasonably good
match to the observed profile variations, so we suspect that 
X-ray photoionization dominates much of the zone between the black 
hole and the facing hemisphere of the supergiant.  
A similar degree of wind ionization probably also exists in 
the X-ray low/hard state since similar orbital variations in 
wind line strength are found in {\it IUE} low dispersion 
spectra made during the X-ray low/hard state.  

Second, the {\it HST} spectra suggest that stellar wind gas emanating from 
parts of the photosphere facing the X-ray source attains only a small velocity
before becoming photoionized.  For example, the highest optical depth 
wind feature, \ion{C}{4} $\lambda\lambda 1548, 1550$, shows only a 
very modest P~Cygni absorption core at phase $\phi=0.5$ (see Fig.~11) 
that extends blueward no more than about $-400$ km~s$^{-1}$ 
(and it is possible that this small component results from a minor 
part of shadow wind projected against supergiant at $\phi=0.5$; see Fig.~8). 
This very low wind speed is probably less than 
the stellar escape velocity ($\sim 700$ km~s$^{-1}$ near the poles). 

These results from the UV wind lines indicate that very little 
mass loss is occurring by a radiatively driven wind for surface 
regions that are exposed to the X-ray source.  The fact that the 
wind features appear similarly weak in {\it IUE} spectra obtained in the 
low/hard state suggests that a spherical, radiatively driven 
wind from the hemisphere of the supergiant facing the black hole 
is probably always weak or absent, and thus, accretion 
from a spherically symmetric wind must play a minor role 
in feeding the black hole in Cyg~X-1. 
 
On the other hand, we found that the emission equivalent width of 
the \ion{He}{2} $\lambda 4686$ line is consistently strong 
between X-ray states and over the available record of observation. 
The orbital phase variations of this spectral feature \citep{gie86b,nin87}
are successfully matched by models of emission from an
enhanced density and slower gas outflow region between the 
stars that is expected for a focused stellar wind \citep{fri82}. 
Thus, while X-ray ionization reduces the wind outflow away from 
the axis joining the stars, the X-ray flux is apparently 
insufficient to stop the outflow in the denser gas of the 
focused wind (the result of both tidal and radiative forces).
Consequently, it is this focused wind component that is 
probably the primary means of mass transfer in Cyg~X-1. 
We caution that the relative constancy of the \ion{He}{2} $\lambda 4686$ 
emission flux does not necessarily imply that the mass loss
rate in the focused wind is also steady.  For example, the 
increased X-ray photoionization during the high/soft state 
may lead to an increase in \ion{He}{2} $\lambda 4686$ emission
(see Fig.~2 in \citealt{gie86b}), so that a lower mass loss 
rate but higher ionization fraction might result in the same 
amount of observed emission.  However, the presence of the 
\ion{He}{2} $\lambda 4686$ emission in both X-ray states 
indicates that focused wind mass loss always occurs at some level. 
 
Finally, we confirm that the H$\alpha$ P~Cygni line forms mainly in 
the base of the stellar wind of the supergiant since we observe 
that H$\alpha$ follows the orbital velocity curve of the supergiant 
(Fig.~1).  The new observations are consistent with earlier results
\citep{gie03,tar03} in demonstrating that the H$\alpha$ emission 
strength is generally weaker in the high/soft X-ray state.  Photoionization 
and heating may extend down to atmospheric levels where the gas densities 
are sufficient to create H$\alpha$ emission, so that the reduction in 
H$\alpha$ strength in the X-ray high/soft state may partially result  
from photoionization related processes.  However, \citet{gie03} 
showed that H$\alpha$ emission variability was present in those
Doppler shifted parts of the profile corresponding to the X-ray 
shadow hemisphere of the supergiant (see their Fig.~15), so part of the 
H$\alpha$ variations must be related to gas density variations at the base 
of the stellar wind.  Thus, the observed H$\alpha$ variations suggest 
that the high/soft X-ray state occurs when the global, radiatively driven 
part of the wind is weaker.  Long term, quasi-cyclic variations in 
wind strength are apparently common among hot supergiants \citep{mar05}. 

\citet{gie03} and \citet{tar03} suggested that the variations
in X-ray state are caused by changes in wind velocity due to 
changes in supergiant mass loss rate.  During times when the 
supergiant's wind is denser and the mass loss rate is higher, 
the photoionization region would be more restricted to 
the region closer to the black hole.  Consequently, a radiatively 
driven wind could accelerate to a higher speed before stalling when 
the gas enters the ionization zone, and thus, the faster wind 
would result in a lower black hole accretion rate and X-ray luminosity
(creating the low/hard X-ray state).  Conversely, if the wind 
mass loss rate drops, then the X-ray ionization zone will expand, 
the maximum wind velocity towards the black hole will decline, 
and the net accretion rate will increase (perhaps creating 
the high/soft state; \citealt{ho87}).  This creates a positive feedback 
mechanism that may continue until the wind is ionized all the way down 
to the stellar photosphere facing the X-ray source 
\citep{day93,blo94}. 

If this scenario is correct, then we expect that the outflow 
velocities in the direction of the black hole (as measured in 
blue extent of the P~Cygni lines at phase $\phi=0.5$) will 
be larger than the supergiant escape velocity during the 
X-ray low/hard state.  The superb quality {\it HST}/STIS spectra 
indicate that outflow velocities are too low to launch the  
wind during the X-ray high/soft state. 
Moreover, the low resolution {\it IUE} spectra from the low/hard state 
appear to show a very similar pattern of the loss of the 
P~Cygni absorption at $\phi=0.5$ (Fig.~12 and 13), indicating that 
significant ionization zones still exist in the low/hard state.
Taken at face value, these {\it IUE} results suggest that 
the spherical component of wind outflow towards the black hole
is weak and slow in both X-ray states, so a wind speed modulation is 
probably not the explanation for the accretion variations 
associated with the X-ray states.  We will require new, high 
quality, UV spectroscopy of Cyg~X-1 during the low/hard state 
in order to make a definitive test of this idea. 

The radiatively driven wind of the supergiant leads to effective 
mass loss only in the X-ray shadowed hemisphere and in the 
focused wind between the stars in Cyg~X-1.  The outflow in the 
shadow wind region will experience a Coriolis deflection, so 
that the trailing regions of the shadow wind will eventually 
enter the zone of X-ray illumination \citep{blo94}.  
Once photoionized, this gas will stall with the loss of the
important ions for radiative acceleration, and some of this slower  
gas may extend around the orbital plane to the vicinity of 
the black hole.  Although this deflected wind gas is 
probably not a major accretion source \citep{blo94}, 
it may affect the accretion dynamics of  
the focused wind.  For example, when the shadow wind mass loss
rate is high (times of strong H$\alpha$ emission), the resulting 
stalled wind component will create a higher ambient gas density 
on the leading side of the zone surrounding the black hole.  
The focused wind flow will make a trajectory towards the following 
side of the black hole, and while gas passing closer to the 
black hole will merge into an accretion disk, gas further out 
will tend to move past the black hole before turning into 
the outskirts of the disk.  The presence of the stalled gas 
on the leading side may deflect away this outer, lower density part of 
the flow and effectively inhibit gas accretion from the focused wind.
The subsequent reduction in gas accretion by the black 
hole may correspond to the conditions required to produce 
the low/hard X-ray state, while conversely a reduction in the 
stalled gas from the shadow wind may promote mass accretion and
produce the high/soft state \citep{bro99,don02,mcc06}.  
Clearly, new hydrodynamical simulations are needed to test whether 
the stalled wind component is sufficient to alter the accretion of gas 
from the focused wind and create the environments needed for the 
X-ray transitions. 


\acknowledgments

We thank the staffs of the David Dunlap Observatory,
Kitt Peak National Observatory, Dominion Astrophysical Observatory,
Ond\v{r}ejov Observatory, and the Space Telescope Science Institute 
(STScI) for their support in obtaining these observations.  
The KPNO spectra supporting the second {\it HST} run were 
obtained with the assistance of participants in the NOAO
Teacher Leaders in Research Based Science Education program, including 
Joan Kadaras (Westford Academy, Westford, MA), Steve Harness (Kingsburg Joint
Union High School, Kingsburg, CA), Elba Sepulveda (CROEM, Mayaguez, PR), and
Dwight Taylor (Goldenview Middle School, Anchorage, AK).
We also thank Saku Vrtilek and Bram Boroson for helpful comments, 
and we are especially grateful to an anonymous referee 
whose report was pivotal to our discussion of the results.  
Support for {\it HST} proposal number GO-9840 was provided by NASA through
a grant from the Space Telescope Science Institute,
which is operated by the Association of Universities for Research
in Astronomy, Incorporated, under NASA contract NAS5-26555.
The X-ray results were provided by the ASM/RXTE teams at 
MIT and at the RXTE SOF and GOF at NASA's GSFC. 
The {\it IUE} data presented in this paper were obtained from 
the Multimission Archive at the Space Telescope Science 
Institute (MAST).  Support for MAST for non-HST data is 
provided by the NASA Office of Space Science via grant 
NAG5-7584 and by other grants and contracts.
Bolton's research is partially supported by a Natural Sciences
and Engineering Research Council of Canada (NSERC) Discovery Grant.
Hadrava's research is funded under grant projects GA\v{C}R 202/06/0041 and LC06014.
Herrero thanks the Spanish MEC for support under project AYA 2007-67456-C02-01.
This work was also supported by the National Science 
Foundation under grants AST-0205297, AST-0506573, and AST-0606861.
Institutional support has been provided from the GSU College
of Arts and Sciences and from the Research Program Enhancement
fund of the Board of Regents of the University System of Georgia,
administered through the GSU Office of the Vice President
for Research.  We are grateful for all this support. 




\clearpage
\begin{deluxetable}{cccccl}
\tablewidth{0pc}
\tabletypesize{\scriptsize}
\rotate
\tablenum{1}
\tablecaption{Journal of Spectroscopy \label{tab1}}
\tablehead{
\colhead{Run} &
\colhead{Dates} &
\colhead{Range} &
\colhead{Resolving Power} &
\colhead{} &
\colhead{Observatory/Telescope/} \\
\colhead{Number} &
\colhead{(HJD-2,450,000)} &
\colhead{(\AA)} &
\colhead{($\lambda/\triangle\lambda$)} &
\colhead{$N$} &
\colhead{Spec., Grating/Detector}}
\startdata
1\dotfill &2419.8 -- 2828.7 & 6510 -- 6710 &\phn8900 &   70 & DDO/1.88m/Cass., 1800 g mm$^{-1}$/Thomson $1024\times1024$ \\
2\dotfill &2448.7 -- 2453.9 & 6530 -- 6710 &\phn7400 &   15 & KPNO/0.9m/Coud\'{e}, B (order 2)/TI5 \\
3\dotfill &2825.8 -- 2828.9 & 6320 -- 8970 &\phn2720 &\phn6 & KPNO/0.9m/Coud\'{e}, RC400 (order 1)/F3KB \\
4\dotfill &2826.8 -- 2827.9 & 6353 -- 6756 &\phn6900 &   10 & DAO/1.85m/Cass., 21121R/SITe-2 \\
5\dotfill &2730.7 -- 2904.5 & 6258 -- 6770 &   10900 &   24 & Ond\v{r}ejov/2m/Coud\'{e}, 700mm/SITe $2000\times800$ \\
6\dotfill &2450.3 -- 2453.3 & 1150 -- 1740 &   14500 &\phn4 & HST/2.4m/STIS, G140M/FUV-MAMA \\
7\dotfill &2825.7 -- 2827.8 & 1150 -- 1740 &   14500 &\phn4 & HST/2.4m/STIS, G140M/FUV-MAMA \\
8\dotfill &2448.7 -- 2454.0 & 4624 -- 4740 &   13000 &   14 & KPNO/0.9m/Coud\'{e}, B (order 3)/TI5 \\
9\dotfill &2824.9 -- 2826.8 & 3759 -- 5086 &\phn2990 &\phn3 & KPNO/0.9m/Coud\'{e}, RC400 (order 2)/F3KB \\
10\dotfill&2912.8 -- 2915.8 & 4182 -- 4942 &\phn5700 &\phn5 & KPNO/4m/RC Spec., BL380 (order 2)/T2KB \\
\enddata
\end{deluxetable}

\begin{deluxetable}{lcccccc}
\tabletypesize{\scriptsize}
\tablewidth{0pt}
\tablenum{2}
\tablecaption{Radial Velocity and H$\alpha$ Equivalent Width Measurements \label{tab2}}
\tablehead{
\colhead{HJD}             &
\colhead{Orbital}         &
\colhead{$V_r$}           &
\colhead{$\triangle V_r$} &
\colhead{$(O-C)$}         &
\colhead{$W_\lambda$(H$\alpha$)}     &
\colhead{Run}             \\
\colhead{(-2,450,000)}    &
\colhead{Phase}           &
\colhead{(km s$^{-1}$)}   &
\colhead{(km s$^{-1}$)}   &
\colhead{(km s$^{-1}$)}   &
\colhead{(\AA )}          &
\colhead{Number} }
\scriptsize
\startdata
2419.847\dotfill & 0.115 & \phs 45.4 & \phn  3.1 & \phn \phs 2.2 &   $-$0.38 & 1 \\
2426.848\dotfill & 0.365 & \phs 55.6 & \phn  4.5 & \phn \phs 6.1 &   $-$0.54 & 1 \\
2427.841\dotfill & 0.543 &   $-$22.9 & \phn  3.2 & \phn \phs 1.6 &   $-$0.75 & 1 \\
2429.605\dotfill & 0.858 &   $-$65.3 & \phn  3.1 & \phn   $-$3.3 &   $-$0.02 & 1 \\
2443.803\dotfill & 0.393 & \phs 42.4 & \phn  3.2 & \phn \phs 2.1 &   $-$0.10 & 1 \\
2445.846\dotfill & 0.758 &   $-$88.6 & \phn  3.1 &       $-$10.6 &   $-$0.07 & 1 \\
2446.804\dotfill & 0.929 &   $-$42.0 & \phn  3.1 & \phn   $-$5.4 &   $-$0.10 & 1 \\
2448.694\dotfill & 0.267 & \phs 68.4 & \phn  1.3 & \phn \phs 1.0 &  \phs0.01 & 2 \\
2448.850\dotfill & 0.294 & \phs 67.5 & \phn  3.1 & \phn \phs 2.5 &  \phs0.10 & 1 \\
2448.954\dotfill & 0.313 & \phs 66.5 & \phn  1.1 & \phn \phs 4.3 &   $-$0.22 & 2 \\
2449.683\dotfill & 0.443 & \phs 18.2 & \phn  1.8 & \phn   $-$2.2 &   $-$0.04 & 2 \\
2449.790\dotfill & 0.462 &\phn\phs7.7& \phn  3.3 & \phn   $-$4.3 &   $-$0.15 & 2 \\
2449.932\dotfill & 0.488 &\phn$-$8.9 & \phn  1.7 & \phn   $-$9.5 &   $-$0.48 & 2 \\
2450.272\dotfill & 0.548 &   $-$27.5 & \phn  3.0 & \phn   $-$0.6 &  \nodata  & 6 \\
2450.335\dotfill & 0.560 &   $-$31.6 & \phn  3.0 & \phn \phs 0.2 &  \nodata  & 6 \\
2450.637\dotfill & 0.614 &   $-$45.7 & \phn  3.2 & \phn \phs 7.2 &   $-$0.43 & 1 \\
2450.667\dotfill & 0.619 &   $-$49.2 & \phn  1.8 & \phn \phs 5.5 &   $-$0.60 & 2 \\
2450.715\dotfill & 0.628 &   $-$60.3 & \phn  3.5 & \phn   $-$2.8 &   $-$0.56 & 1 \\
2450.760\dotfill & 0.636 &   $-$60.2 & \phn  1.3 & \phn   $-$0.2 &   $-$0.72 & 2 \\
2450.792\dotfill & 0.641 &   $-$53.6 & \phn  3.1 & \phn \phs 8.1 &   $-$0.38 & 1 \\
2450.812\dotfill & 0.645 &   $-$66.3 & \phn  3.1 & \phn   $-$3.6 &   $-$0.83 & 1 \\
2450.912\dotfill & 0.663 &   $-$66.6 & \phn  1.6 & \phn \phs 0.8 &   $-$0.74 & 2 \\
2451.603\dotfill & 0.786 &   $-$79.6 & \phn  3.1 & \phn   $-$3.4 &   $-$0.69 & 1 \\
2451.642\dotfill & 0.793 &   $-$82.9 & \phn  3.1 & \phn   $-$7.4 &   $-$0.33 & 1 \\
2451.642\dotfill & 0.793 &   $-$82.3 & \phn  3.1 & \phn   $-$6.9 &   $-$0.35 & 1 \\
2451.686\dotfill & 0.801 &   $-$76.8 & \phn  3.2 & \phn   $-$2.5 &   $-$0.54 & 1 \\
2451.701\dotfill & 0.803 &   $-$77.0 & \phn  1.6 & \phn   $-$3.0 &   $-$0.45 & 2 \\
2451.709\dotfill & 0.805 &   $-$79.7 & \phn  3.2 & \phn   $-$6.0 &   $-$0.73 & 1 \\
2451.737\dotfill & 0.810 &   $-$76.0 & \phn  3.1 & \phn   $-$3.1 &   $-$0.41 & 1 \\
2452.711\dotfill & 0.984 &   $-$12.9 & \phn  3.1 & \phn   $-$0.4 &  \phs0.11 & 1 \\
2452.733\dotfill & 0.988 &   $-$10.9 & \phn  3.1 & \phn   $-$0.2 &   $-$0.04 & 1 \\
2452.734\dotfill & 0.988 &\phn$-$9.1 & \phn  1.6 & \phn \phs 1.4 &  \phs0.04 & 2 \\
2452.772\dotfill & 0.995 &\phn$-$5.5 & \phn  3.2 & \phn \phs 2.0 &  \phs0.13 & 1 \\
2452.795\dotfill & 0.999 &\phn$-$5.3 & \phn  3.2 & \phn \phs 0.3 &   $-$0.34 & 1 \\
2452.836\dotfill & 0.006 &\phn$-$1.2 & \phn  1.3 & \phn \phs 1.1 &  \phs0.16 & 2 \\
2452.962\dotfill & 0.029 & \phs 12.4 & \phn  1.6 & \phn \phs 4.4 &  \phs0.10 & 2 \\
2453.210\dotfill & 0.073 & \phs 34.0 & \phn  3.0 & \phn \phs 6.9 &  \nodata  & 6 \\
2453.272\dotfill & 0.084 & \phs 34.5 & \phn  3.0 & \phn \phs 2.8 &  \nodata  & 6 \\
2453.664\dotfill & 0.154 & \phs 59.2 & \phn  1.5 & \phn \phs 4.2 &   $-$0.21 & 2 \\
2453.744\dotfill & 0.168 & \phs 57.4 & \phn  3.1 & \phn   $-$1.1 &   $-$0.23 & 1 \\
2453.754\dotfill & 0.170 & \phs 56.7 & \phn  1.2 & \phn   $-$2.1 &   $-$0.22 & 2 \\
2453.766\dotfill & 0.172 & \phs 53.3 & \phn  3.1 & \phn   $-$6.0 &   $-$0.22 & 1 \\
2453.806\dotfill & 0.180 & \phs 64.5 & \phn  3.2 & \phn \phs 3.7 &   $-$0.18 & 1 \\
2453.828\dotfill & 0.183 & \phs 59.8 & \phn  3.1 & \phn   $-$1.8 &   $-$0.13 & 1 \\
2453.845\dotfill & 0.186 & \phs 53.1 & \phn  3.1 & \phn   $-$9.0 &   $-$0.66 & 1 \\
2453.934\dotfill & 0.202 & \phs 59.9 & \phn  1.3 & \phn   $-$4.7 &   $-$0.09 & 2 \\
2454.635\dotfill & 0.327 & \phs 57.3 & \phn  3.1 & \phn   $-$2.1 &   $-$0.09 & 1 \\
2454.660\dotfill & 0.332 & \phs 52.6 & \phn  3.1 & \phn   $-$5.8 &  \phs0.09 & 1 \\
2454.682\dotfill & 0.336 & \phs 53.4 & \phn  3.1 & \phn   $-$4.0 &  \phs0.12 & 1 \\
2454.705\dotfill & 0.340 & \phs 49.5 & \phn  3.1 & \phn   $-$7.0 &   $-$0.07 & 1 \\
2454.728\dotfill & 0.344 & \phs 44.7 & \phn  3.1 &       $-$10.8 &   $-$0.05 & 1 \\
2454.750\dotfill & 0.348 & \phs 44.8 & \phn  3.1 & \phn   $-$9.6 &   $-$0.17 & 1 \\
2454.773\dotfill & 0.352 & \phs 43.5 & \phn  3.1 & \phn   $-$9.8 &   $-$0.22 & 1 \\
2454.796\dotfill & 0.356 & \phs 39.6 & \phn  3.1 &       $-$12.6 &   $-$0.07 & 1 \\
2454.818\dotfill & 0.360 & \phs 41.2 & \phn  3.1 & \phn   $-$9.9 &   $-$0.04 & 1 \\
2454.841\dotfill & 0.364 & \phs 43.6 & \phn  3.1 & \phn   $-$6.2 &   $-$0.03 & 1 \\
2454.859\dotfill & 0.368 & \phs 44.2 & \phn  3.1 & \phn   $-$4.6 &   $-$0.55 & 1 \\
2463.626\dotfill & 0.933 &   $-$43.2 & \phn  3.2 & \phn   $-$8.3 &   $-$0.15 & 1 \\
2479.857\dotfill & 0.832 &   $-$64.9 & \phn  3.1 & \phn \phs 3.8 &   $-$0.13 & 1 \\
2519.667\dotfill & 0.941 &   $-$31.2 & \phn  3.1 & \phn \phs 0.5 &   $-$0.30 & 1 \\
2524.692\dotfill & 0.838 &   $-$70.8 & \phn  3.1 & \phn   $-$3.6 &   $-$0.28 & 1 \\
2530.586\dotfill & 0.891 &   $-$47.5 & \phn  3.1 & \phn \phs 3.9 &   $-$0.13 & 1 \\
2554.578\dotfill & 0.175 & \phs 63.8 & \phn  3.1 & \phn \phs 3.9 &   $-$0.08 & 1 \\
2571.581\dotfill & 0.211 & \phs 65.1 & \phn  3.1 & \phn   $-$0.6 &   $-$0.77 & 1 \\
2576.659\dotfill & 0.118 & \phs 47.9 & \phn  3.1 & \phn \phs 3.8 &   $-$0.29 & 1 \\
2578.510\dotfill & 0.449 & \phs 12.8 & \phn  3.1 & \phn   $-$5.1 &   $-$0.39 & 1 \\
2580.562\dotfill & 0.815 &   $-$70.5 & \phn  3.1 & \phn \phs 1.6 &   $-$0.73 & 1 \\
2592.565\dotfill & 0.959 &   $-$29.0 & \phn  3.1 & \phn   $-$5.1 &   $-$1.05 & 1 \\
2611.473\dotfill & 0.335 & \phs 57.2 & \phn  3.1 & \phn   $-$0.5 &   $-$0.57 & 1 \\
2613.498\dotfill & 0.697 &   $-$78.6 & \phn  3.1 & \phn   $-$4.5 &   $-$0.83 & 1 \\
2619.458\dotfill & 0.761 &   $-$79.6 & \phn  3.1 & \phn   $-$1.7 &   $-$0.85 & 1 \\
2625.476\dotfill & 0.836 &   $-$75.7 & \phn  3.3 & \phn   $-$8.0 &   $-$0.90 & 1 \\
2626.457\dotfill & 0.011 &\phn$-$4.5 & \phn  3.1 & \phn   $-$4.4 &   $-$0.57 & 1 \\
2638.476\dotfill & 0.157 & \phs 60.8 & \phn  3.1 & \phn \phs 5.0 &   $-$0.60 & 1 \\
2646.474\dotfill & 0.585 &   $-$31.6 & \phn  3.1 &      \phs10.8 &   $-$1.05 & 1 \\
2709.867\dotfill & 0.906 &   $-$44.6 & \phn  3.3 & \phn \phs 1.2 &   $-$0.70 & 1 \\
2730.654\dotfill & 0.618 &   $-$40.0 & \phn  5.0 &      \phs14.4 &   $-$1.21 & 5 \\
2730.755\dotfill & 0.636 &   $-$64.4 & \phn  4.2 & \phn   $-$4.2 &   $-$1.82 & 1 \\
2744.818\dotfill & 0.148 & \phs 56.7 & \phn  3.1 & \phn \phs 3.4 &   $-$1.25 & 1 \\
2746.612\dotfill & 0.468 & \phs 14.6 & \phn  3.0 & \phn \phs 5.2 &   $-$1.34 & 5 \\
2762.871\dotfill & 0.371 & \phs 48.3 & \phn  3.1 & \phn \phs 0.7 &   $-$0.64 & 1 \\
2805.784\dotfill & 0.035 &\phn\phs2.3& \phn  3.1 & \phn   $-$8.3 &   $-$0.35 & 1 \\
2813.681\dotfill & 0.445 & \phs 18.4 & \phn  3.1 & \phn   $-$1.2 &   $-$0.53 & 1 \\
2824.647\dotfill & 0.403 & \nodata   & \nodata   & \nodata       &   $-$0.29 & 1 \\
2825.702\dotfill & 0.592 &   $-$31.8 & \phn  3.0 &      \phs13.0 &  \nodata  & 7 \\
2825.765\dotfill & 0.603 &   $-$41.1 & \phn  3.0 & \phn \phs 7.9 &  \nodata  & 7 \\
2825.839\dotfill & 0.616 &   $-$44.3 & \phn  4.6 & \phn \phs 9.4 &   $-$0.40 & 3 \\
2825.844\dotfill & 0.617 &   $-$43.3 &      10.7 &      \phs10.7 &   $-$0.36 & 3 \\
2826.817\dotfill & 0.791 &   $-$65.7 & \phn  3.1 &      \phs10.1 &   $-$0.34 & 1 \\
2826.832\dotfill & 0.793 &   $-$49.9 & \phn  8.0 &      \phs25.5 &   $-$0.44 & 4 \\
2826.860\dotfill & 0.798 &   $-$69.2 & \phn  5.4 & \phn \phs 5.6 &   $-$0.29 & 4 \\
2826.885\dotfill & 0.803 &   $-$69.4 & \phn  5.2 & \phn \phs 4.7 &   $-$0.36 & 4 \\
2826.909\dotfill & 0.807 &   $-$67.3 & \phn  5.4 & \phn \phs 6.2 &   $-$0.37 & 4 \\
2826.933\dotfill & 0.811 &   $-$62.6 & \phn  5.2 &      \phs10.1 &   $-$0.43 & 4 \\
2827.702\dotfill & 0.949 &   $-$27.9 & \phn  3.0 & \phn \phs 0.3 &  \nodata  & 7 \\
2827.765\dotfill & 0.960 &   $-$26.4 & \phn  3.0 & \phn   $-$3.0 &  \nodata  & 7 \\
2827.800\dotfill & 0.966 &   $-$23.7 & \phn  6.8 & \phn   $-$3.2 &   $-$0.44 & 3 \\
2827.812\dotfill & 0.968 &   $-$23.7 & \phn  7.2 & \phn   $-$4.1 &   $-$0.47 & 3 \\
2827.812\dotfill & 0.968 &   $-$23.0 & \phn  5.2 & \phn   $-$3.5 &   $-$0.60 & 4 \\
2827.836\dotfill & 0.973 &   $-$21.7 & \phn  5.2 & \phn   $-$4.1 &   $-$0.55 & 4 \\
2827.882\dotfill & 0.981 &   $-$13.5 & \phn  5.3 & \phn \phs 0.4 &   $-$0.68 & 4 \\
2827.905\dotfill & 0.985 &   $-$14.0 & \phn  5.2 & \phn   $-$2.0 &   $-$0.60 & 4 \\
2827.929\dotfill & 0.989 &\phn$-$7.3 & \phn  5.2 & \phn \phs 2.8 &   $-$0.79 & 4 \\
2828.594\dotfill & 0.108 & \phs 40.7 & \phn  3.1 & \phn \phs 0.1 &   $-$0.16 & 1 \\
2828.657\dotfill & 0.119 & \phs 45.3 & \phn  3.1 & \phn \phs 0.7 &  \phs0.08 & 1 \\
2828.680\dotfill & 0.123 & \phs 45.6 & \phn  3.1 & \phn   $-$0.3 &   $-$0.09 & 1 \\
2828.702\dotfill & 0.127 & \phs 46.4 & \phn  3.1 & \phn   $-$0.8 &   $-$0.04 & 1 \\
2828.725\dotfill & 0.131 & \phs 45.4 & \phn  3.1 & \phn   $-$3.0 &   $-$0.13 & 1 \\
2828.747\dotfill & 0.135 & \phs 50.0 & \phn  3.1 & \phn \phs 0.3 &   $-$0.03 & 1 \\
2828.885\dotfill & 0.160 & \phs 57.5 & \phn  5.5 & \phn \phs 1.1 &   $-$0.20 & 3 \\
2828.892\dotfill & 0.161 & \phs 55.5 & \phn  4.6 & \phn   $-$1.3 &   $-$0.17 & 3 \\
2835.440\dotfill & 0.330 & \phs 61.4 & \phn  2.9 & \phn \phs 2.7 &  \phs0.00 & 5 \\
2840.510\dotfill & 0.236 & \nodata   & \nodata   & \nodata       &   $-$0.09 & 5 \\
2846.494\dotfill & 0.305 & \phs 64.8 & \phn  2.9 & \phn \phs 1.2 &   $-$0.53 & 5 \\
2857.443\dotfill & 0.260 & \phs 71.3 & \phn  2.9 & \phn \phs 3.6 &   $-$0.09 & 5 \\
2857.471\dotfill & 0.265 & \phs 69.7 & \phn  2.9 & \phn \phs 2.1 &   $-$0.24 & 5 \\
2859.510\dotfill & 0.629 &   $-$72.4 & \phn  2.9 &       $-$14.4 &   $-$0.52 & 5 \\
2860.535\dotfill & 0.812 &   $-$72.9 & \phn  2.9 & \phn   $-$0.2 &   $-$0.48 & 5 \\
2860.558\dotfill & 0.816 &   $-$71.8 & \phn  2.9 & \phn \phs 0.1 &   $-$0.47 & 5 \\
2861.539\dotfill & 0.991 &\phn$-$7.9 & \phn  2.9 & \phn \phs 1.3 &   $-$0.31 & 5 \\
2862.547\dotfill & 0.171 & \phs 61.6 & \phn  3.2 & \phn \phs 2.5 &   $-$0.35 & 5 \\
2862.572\dotfill & 0.176 & \phs 58.2 & \phn  2.9 & \phn   $-$1.8 &   $-$0.35 & 5 \\
2874.472\dotfill & 0.301 & \phs 72.1 & \phn  2.9 & \phn \phs 7.9 &   $-$0.82 & 5 \\
2874.494\dotfill & 0.305 & \phs 67.6 & \phn  2.9 & \phn \phs 4.1 &   $-$0.76 & 5 \\
2874.516\dotfill & 0.309 & \phs 72.6 & \phn  2.9 & \phn \phs 9.6 &   $-$0.70 & 5 \\
2874.538\dotfill & 0.312 & \phs 70.5 & \phn  2.9 & \phn \phs 8.2 &   $-$0.66 & 5 \\
2878.417\dotfill & 0.005 &\phn\phs5.0& \phn  2.9 & \phn \phs 7.7 &   $-$0.68 & 5 \\
2878.441\dotfill & 0.009 &\phn\phs8.2& \phn  2.9 & \phn \phs 9.0 &   $-$0.54 & 5 \\
2898.466\dotfill & 0.585 &   $-$36.2 & \phn  2.9 & \phn \phs 6.2 &   $-$1.14 & 5 \\
2901.357\dotfill & 0.102 & \phs 38.4 & \phn  3.1 & \phn \phs 0.0 &   $-$1.26 & 5 \\
2903.436\dotfill & 0.473 & \phs 22.2 & \phn  2.9 &      \phs14.9 &   $-$0.96 & 5 \\
2903.475\dotfill & 0.480 & \phs 16.9 & \phn  2.9 &      \phs12.9 &   $-$0.98 & 5 \\
2904.480\dotfill & 0.659 &   $-$55.5 & \phn  2.9 &      \phs11.1 &   $-$1.12 & 5 
\enddata
\end{deluxetable}


\begin{deluxetable}{lccc}
\tablewidth{0pc}
\tablenum{3}
\tablecaption{Circular Orbital Elements\label{tab3}}
\tablehead{
\colhead{Element} &\colhead{Brocksopp et al.\ (1999)} &\colhead{Gies et al.\ (2003)} &\colhead{This Work} }
\startdata
$P$ (d)                 \dotfill & 5.599829 (16)      & 5.599829\tablenotemark{a}& 5.599829\tablenotemark{a} \\
$T$(IC) (HJD-2,400,000) \dotfill & 41,874.707 (9)     & 51,730.449 (8)           & 52,872.788 (9) \\
$K_1$ (km s$^{-1}$)     \dotfill & 74.9 (6)           & 75.6 (7)                 & 73.0 (7) \\
$V_0$ (km s$^{-1}$)     \dotfill & \nodata            & $-$7.0 (5)               & $-$5.1 (5) \\
$\sigma$ (km s$^{-1}$)  \dotfill & \nodata            & 5.3                      & 5.4  \\
\enddata
\tablenotetext{a}{Fixed.}
\tablecomments{Numbers in parentheses give the error in the last digit quoted.}
\end{deluxetable}


\begin{deluxetable}{lccc}
\tablewidth{0pc}
\tablenum{4}
\tablecaption{H$\alpha$ Equivalent Widths and X-ray Flux Counts for the {\it HST} Observation Times\label{tab4}}
\tablehead{
\colhead{HJD}             &
\colhead{Orbital}         &
\colhead{$W_\lambda$(H$\alpha$)}   &
\colhead{Flux (1.5 -- 3 keV)}  \\
\colhead{(-2,450,000)}    &
\colhead{Phase}           &
\colhead{(\AA )}          &
\colhead{(ASM counts)} }
\startdata
2450.3\dotfill & 0.55 & $-$0.45 (15) & 52 (10) \\
2453.2\dotfill & 0.08 & $-$0.02 (10) & 60 (10) \\
2825.7\dotfill & 0.60 & $-$0.39 (10) & 50 (10) \\
2827.7\dotfill & 0.95 & $-$0.44 (10) & 42 (10) \\
\enddata
\tablecomments{Numbers in parentheses give the error in the last digit quoted.}
\end{deluxetable}


\begin{deluxetable}{lccccc}
\tablewidth{0pc}
\tablenum{5}
\tablecaption{\ion{He}{2} $\lambda4686$ Equivalent Widths\label{tab5}}
\tablehead{
\colhead{HJD}             &
\colhead{Orbital}         &
\colhead{$W_\lambda$}     &
\colhead{$\triangle W_\lambda$}     &
\colhead{Run}             &
\colhead{X-ray}           \\
\colhead{(-2,450,000)}    &
\colhead{Phase}           &
\colhead{(\AA )}          &
\colhead{(\AA )}          &
\colhead{Number}          &
\colhead{State} }
\startdata
 52448.748\dotfill & 0.276 & $-0.39$ &  0.03 & \phn8 & high/soft \\
 52448.905\dotfill & 0.304 & $-0.35$ &  0.02 & \phn8 & high/soft \\
 52449.733\dotfill & 0.452 & $-0.21$ &  0.03 & \phn8 & high/soft \\
 52449.885\dotfill & 0.479 & $-0.22$ &  0.02 & \phn8 & high/soft \\
 52449.969\dotfill & 0.494 & $-0.24$ &  0.03 & \phn8 & high/soft \\
 52450.713\dotfill & 0.627 & $-0.53$ &  0.03 & \phn8 & high/soft \\
 52450.855\dotfill & 0.653 & $-0.50$ &  0.02 & \phn8 & high/soft \\
 52450.958\dotfill & 0.671 & $-0.49$ &  0.02 & \phn8 & high/soft \\
 52451.732\dotfill & 0.809 & $-0.56$ &  0.07 & \phn8 & high/soft \\
 52452.796\dotfill & 0.999 & $-0.47$ &  0.08 & \phn8 & high/soft \\
 52452.930\dotfill & 0.023 & $-0.40$ &  0.02 & \phn8 & high/soft \\
 52453.709\dotfill & 0.162 & $-0.49$ &  0.05 & \phn8 & high/soft \\
 52453.889\dotfill & 0.194 & $-0.47$ &  0.03 & \phn8 & high/soft \\
 52453.965\dotfill & 0.208 & $-0.44$ &  0.03 & \phn8 & high/soft \\
 52824.884\dotfill & 0.445 & $-0.38$ &  0.06 & \phn9 & high/soft \\
 52824.891\dotfill & 0.447 & $-0.23$ &  0.06 & \phn9 & high/soft \\
 52826.841\dotfill & 0.795 & $-0.25$ &  0.04 & \phn9 & high/soft \\
 52912.769\dotfill & 0.140 & $-0.46$ &  0.02 &    10 & low/hard  \\
 52912.773\dotfill & 0.140 & $-0.49$ &  0.02 &    10 & low/hard  \\
 52913.768\dotfill & 0.318 & $-0.46$ &  0.02 &    10 & low/hard  \\
 52914.766\dotfill & 0.496 & $-0.23$ &  0.02 &    10 & low/hard  \\
 52915.761\dotfill & 0.674 & $-0.31$ &  0.02 &    10 & low/hard  \\
\enddata
\end{deluxetable}


\begin{deluxetable}{lccccc}
\tablewidth{0pc}
\tablenum{6}
\tablecaption{\ion{He}{2} $\lambda4686$ Equivalent Width Averages\label{tab6}}
\tablehead{
\colhead{Observation Dates}  &
\colhead{$W_\lambda$}        &
\colhead{$\sigma (W_\lambda)$} &
\colhead{ }                  \\
\colhead{(BY)}               &
\colhead{(\AA )}             &
\colhead{(\AA )}             &
\colhead{Source} }
\startdata
1971--1981 \dotfill & $-0.36$ &   0.12  & \citet{gie86a} \\ 
1980--1984 \dotfill & $-0.29$ &   0.09  & \citet{nin87} \\  
1996--1998 \dotfill & $-0.25$ &   0.12  & \citet{bro99}  \\
2002--2003 \dotfill & $-0.39$ &   0.12  & This paper, high/soft state \\
2003       \dotfill & $-0.39$ &   0.11  & This paper, low/hard state  \\
\enddata
\end{deluxetable}


\begin{deluxetable}{lcccccl}
\tablewidth{0pc}
\tablenum{7}
\tablecaption{Mean Flux Across the Wind Lines\label{tab7}}
\tablehead{
\colhead{HJD}             &
\colhead{Orbital}         &
\colhead{Mean Flux}       &
\colhead{Mean Flux}       &
\colhead{}                &
\colhead{}                &
\colhead{}                \\
\colhead{(-2,400,000)}    &
\colhead{Phase}           &
\colhead{(\ion{Si}{4} $\lambda 1400$)}          &
\colhead{(\ion{C}{4} $\lambda 1550$)}          &
\colhead{Source}          &
\colhead{X-ray State}     &
\colhead{Ref.} }
\startdata
43629.222\dotfill & 0.313 & 1.079 & 0.622 & IUE & low  & 1,2,3 \\
43629.917\dotfill & 0.437 & 1.109 & 0.725 & IUE & low  & 1,2,3 \\
43630.898\dotfill & 0.612 & 1.094 & 0.691 & IUE & low  & 1,2,3 \\
43632.884\dotfill & 0.967 & 0.854 & 0.632 & IUE & low  & 1,2,3 \\
43636.069\dotfill & 0.536 & 1.099 & 0.756 & IUE & low  & 1,2,3 \\
43638.907\dotfill & 0.042 & 0.898 & 0.667 & IUE & low  & 1,2,3 \\
43701.758\dotfill & 0.266 & 0.965 & 0.728 & IUE & low  & 1,2,3 \\
43708.857\dotfill & 0.534 & 1.069 & 0.765 & IUE & low  & 1,2,3 \\
43799.446\dotfill & 0.711 & 0.963 & 0.733 & IUE & low  & 1,2,3 \\
43802.485\dotfill & 0.254 & 0.901 & 0.649 & IUE & low  & 1,2,3 \\
43804.629\dotfill & 0.636 & 0.995 & 0.789 & IUE & low  & 1,2,3 \\
43845.978\dotfill & 0.020 & 0.753 & 0.588 & IUE & low  & 1,2,3 \\
43848.020\dotfill & 0.385 & 0.993 & 0.735 & IUE & low  & 1,2,3 \\
43892.220\dotfill & 0.278 & 1.004 & 0.751 & IUE & low  & 1,2,3 \\
43894.744\dotfill & 0.729 & 0.932 & 0.678 & IUE & low  & 1,2,3 \\
43894.811\dotfill & 0.741 & 0.886 & 0.671 & IUE & low  & 1,2,3 \\
43896.419\dotfill & 0.028 & 0.795 & 0.648 & IUE & low  & 1,2,3 \\
44002.919\dotfill & 0.046 & 0.814 & 0.610 & IUE & low  & 1,2,3 \\
44003.118\dotfill & 0.082 & 0.843 & 0.654 & IUE & low  & 1,2,3 \\
44003.259\dotfill & 0.107 & 0.843 & 0.533 & IUE & low  & 1,2,3 \\
44035.011\dotfill & 0.777 & 0.887 & 0.696 & IUE & low  & 1,2,3 \\
44265.582\dotfill & 0.952 & 0.739 & 0.548 & IUE & low  & 1,4 \\
44412.390\dotfill & 0.168 & 0.929 & 0.683 & IUE & high & 5     \\
44416.117\dotfill & 0.834 & 0.817 & 0.577 & IUE & high & 5     \\
44419.259\dotfill & 0.395 & 0.924 & 0.764 & IUE & high & 5     \\
44419.797\dotfill & 0.491 & 1.009 & 0.805 & IUE & high & 5     \\
44422.310\dotfill & 0.940 & 0.840 & 0.678 & IUE & high & 5,6   \\
44423.302\dotfill & 0.117 & 0.790 & 0.584 & IUE & high & 5,6   \\
44425.301\dotfill & 0.474 & 0.993 & 0.746 & IUE & high & 5,6   \\
44426.941\dotfill & 0.767 & 0.962 & 0.757 & IUE & high & 5,6   \\
52450.272\dotfill & 0.548 & 0.947 & 0.714 & HST & high & 7     \\
52450.335\dotfill & 0.560 & 0.947 & 0.722 & HST & high & 7     \\
52453.210\dotfill & 0.073 & 0.752 & 0.515 & HST & high & 7     \\
52453.272\dotfill & 0.084 & 0.761 & 0.520 & HST & high & 7     \\
52825.702\dotfill & 0.592 & 1.030 & 0.743 & HST & high & 7     \\
52825.765\dotfill & 0.603 & 1.014 & 0.725 & HST & high & 7     \\
52827.702\dotfill & 0.949 & 0.804 & 0.551 & HST & high & 7     \\
52827.765\dotfill & 0.960 & 0.787 & 0.530 & HST & high & 7     \\
\enddata
\tablerefs{
1. \citet{lin83};
2. \citet*{kem81};
3. \citet*{pri83};
4. \citet{per86};
5. \citet{oga82};
6. \citet{oda80};
7. This paper.
}
\end{deluxetable}



\clearpage

\clearpage

\input{epsf}

\clearpage

\input{epsf}

\clearpage

\input{epsf}

\clearpage

\input{epsf}

\clearpage

\input{epsf}

\clearpage

\input{epsf}

\clearpage

\input{epsf}

\clearpage

\input{epsf}

\clearpage

\input{epsf}

\clearpage

\input{epsf}

\input{epsf}

\input{epsf}

\input{epsf}

\input{epsf}


\end{document}